\documentclass[aps,prb,twocolumn,floatfix,footinbib,showpacs]{revtex4-1}
\usepackage{graphicx}
\usepackage{amsfonts,amsmath,amssymb}
\usepackage{amsthm}
\usepackage{dsfont,bm}
\usepackage{color}
\usepackage{soul,xcolor}
\setstcolor{blue}
\usepackage{amsbsy}
\usepackage[colorlinks=true,linkcolor=blue,pagecolor=blue,filecolor=blue,menucolor=blue,urlcolor=blue,citecolor=blue,anchorcolor=blue]{hyperref}

\usepackage{mathbbol}
\usepackage{lineno}

\begin{document}


\title{Machine-learned model Hamiltonian and strength of spin-orbit interaction in strained $\rm Mg_2X~(X=Si,Ge,Sn,Pb)$ }

\author{Mohammad Alidoust}
\affiliation{Department of Physics, Norwegian University of Science and Technology, N-7491 Trondheim, Norway}
\author{Erling Rothmund}
\affiliation{Department of Physics, Norwegian University of Science and Technology, N-7491 Trondheim, Norway}
\author{Jaakko Akola} 
\affiliation{Department of Physics, Norwegian University of Science and Technology, N-7491 Trondheim, Norway}
\affiliation{Computational Physics Laboratory, Faculty of Natural Sciences, Tampere University, FI-33101 Tampere, Finland}

\begin{abstract}
Machine-learned multi-orbital tight-binding (MMTB) Hamiltonian models have been developed to describe the electronic characteristics of intermetallic compounds $\rm Mg_2Si, Mg_2Ge, Mg_2Sn$, and $\rm Mg_2Pb$ subject to strain. The MMTB models incorporate spin-orbital mediated interactions and they are calibrated to the electronic band structures calculated via density functional theory (DFT) by a massively parallelized multi-dimensional Monte-Carlo search algorithm. The results show that a machine-learned five-band tight-binding model reproduces the key aspects of the valence band structures in the entire Brillouin zone. The five-band model reveals that compressive strain localizes the contribution of the $3s$ orbital of $\rm Mg$ to the conduction bands and the outer shell $p$ orbitals of $\rm X~(X=Si,Ge,Sn,Pb)$ to the valence bands. In contrast, tensile strain has a reversed effect as it weakens the contribution of the $3s$ orbital of $\rm Mg$ and the outer shell $p$ orbitals of $\rm X$ to the conduction bands and valence bands, respectively. The $\pi$ bonding in the $\rm Mg_2X$ compounds is negligible compared to the $\sigma$ bonding components, which follow the hierarchy $|\sigma_{sp}|>|\sigma_{pp}|>|\sigma_{ss}|$, and the largest variation against strain belongs to $\sigma_{pp}$. The five-band model allows for estimating the strength of spin-orbit coupling (SOC) in $\rm Mg_2X$ and obtaining its dependence on the atomic number of $\rm X$ and strain. Further, the band structure calculations demonstrate a significant band gap tuning and band splitting due to strain. A compressive strain of $-10\%$ can open a band gap at the $\Gamma$ point in metallic $\rm Mg_2Pb$, whereas a tensile strain of $+10\%$ closes the semiconducting band gap of $\rm Mg_2Si$. A tensile strain of  $+5\%$ removes the three-fold degeneracy of valence bands at the $\Gamma$ point in semiconducting $\rm Mg_2Ge$. The presented MMTB models can be extended for various materials and simulations (band structure, transport, classical molecular dynamics), and the obtained results can help in designing devices made of $\rm Mg_2X$.
\end{abstract}

\date{\today}

\maketitle

\section{introduction}

Intermetallic compounds $\rm Mg_2X~(X=Si,Ge,Sn)$ are mainly considered to be semiconductors and their addition to, e.g., metallic matrices can form solid solutes and grains. These $\rm Mg_2X$ solid solutes can desirably change the electronic, mechanical, and macroscopic properties of these metallic matrices.\cite{G.H.Grosch,A.Nozariasb,H.Kamila1,Z.Zhou,S.Muthiah,H.Kamila2} Furthermore, $\rm Mg_2X~(X=Si,Ge,Sn)$ are chemically and thermally stable, nontoxic, resistive against oxidation, economically inexpensive, environmentally friendly, relatively light weight, and therefore have a great potential for large-scale production. \cite{A.Nozariasb,B.Ryu,G.Shi} More importantly, they possess a high figure of merit (thermoelectric performance in a device form) and can serve as good solid-state thermoelectric materials for converting waste heat to electricity. These excellent characteristics have fuelled robust effort to explore various aspects of $\rm Mg_2X$ and find ways to improve and optimize their favorable performance. For instance, one main approach followed to manipulate the electrical transport of these compounds was by intercalation with differing elements. \cite{S.Kim,P.Zwolenski1,P.Zwolenski2,B.Ryu,H.Kamila1,Z.Zhou,H.Kamila22,J.deBoor,J.M.Guerra,J.J.Pulikkotil,X.J.Tan,G.Jiang,W.Liu,J.Mao,G.Shi,M.Yasseri3} It was found that the intercalation can result in nonlinearity in the band gap, enhanced spin-orbit interaction (SOI), and band splitting. The controllable and efficient thermoelectric materials are promising in designing micro-scale self-powered sensors, solar thermal elements, and waste heat recovery devices. \cite{X.Hu,X.Tang,D.Kraemer,X.Shi,N.Espinosa,H.Zhu,K.Kutorasinski,V.Val.Sobolev,O.C.Yelgel}

There has been an intensive theoretical effort to study these compounds, mainly using density functional theory (DFT).\cite{G.H.Grosch,B.Arnaud,H.Mizoguchi,Y.ElAhmar,J.M.Guerra,T.Fan,J.J.Pulikkotil,H.Balout,G.Murtaza,L.Bao2,G.Bai,L.Bao1} The theoretical studies can provide deep and independent insight into the physics of these compositions and shed light on delicate aspects. For example, one can attain a better view over the physical mechanisms and possibly harnessing them in the future experiments, to create new opportunities, achieve more reliable analysis, and avoid introducing detrimental effects during the implementation process. Additionally, there are multiple works discussing various aspects of these compounds by using an effective single-band parabolic approximation. \cite{H.Kamila2,J.M.Guerra,G.Shi} The common assumption of these works is the homogeneity of the alloys and generalization of the rigid-band structure approach to obtain the band structure of the intermediate alloy systems. 

Motivated by the above, we have constructed machine-learned multi-orbital tight-binding (MMTB) models, accounting for the $\{ s,p,d,s^*\}$ orbitals and spin-orbit interaction (SOI). The parameters of the models are determined through optimizing it to DFT band structures by a massively parallelized multi-dimensional Monte-Carlo search algorithm. Our investigations demonstrate that a simple machine-learned five-band tight-binding (TB) model is sufficient to describe the electronic properties of strained $\rm Mg_2X (X=Si,Ge,Sn,Pb)$ close to the Fermi level. The results of the model show that compressive (tensile) strain enhances (suppresses) the SOI strength in $\rm Mg_2X$. The $\rm Mg$-$\rm X$ interactions were found to dominate over $\rm Mg$-$\rm Mg$ and $\rm X$-$\rm X$ counterparts. The $\pi$ bonding is negligible whereas the dominate variation against strain appears in $\sigma_{pp}$. We have found that a tensile strain on the order of $\varepsilon=5\%$ eliminates the three-fold degeneracy of the valence bands at the $\Gamma$ point in $\rm Mg_2Ge$. A two-fold band degeneracy, however, remains intact throughout $\varepsilon\in[-10\%,+10\%]$ in $\rm Mg_2Si$. Our results illustrate that the band gap of $\rm Mg_2X$ is highly tunable through a uniform strain so that, despite their unstrained characteristics, the $\rm Mg_2X$ compounds can acquire semiconductor (metallic) characteristics. These findings point into an excellent opportunity for band engineering of $\rm Mg_2X$-based materials \cite{H.Zhu} and can be helpful when analyzing and designing devices made of $\rm Mg_2X$. 

Moreover, unlike the simple single-band parabolic models, our machine-learned five-band TB model is able to capture key features of the DFT valence band structure of strained $\rm Mg_2X~(X=Si,Ge,Sn,Pb)$ within the entire Brillouin zone with significantly reduced computational complications and lower cost than DFT. Therefore, it can be employed to perform large-scale molecular dynamics simulations and quantum transport studies as further improvements are desired for a practical implementation of $\rm Mg_2X$-based thermoelectric generators, which are of fundamental importance from the engineering and technical points of view. Methodologically, our approach is transferable for $\rm Mg_2X$-related intercalation compounds and it provides a basis for developing more realistic TB models than the single-band parabolic models. We remark that the DFT method itself is applicable for systems with few hundreds of atoms and once different elements are included, the calculations for searching favorable compositions become computationally very challenging if not impossible.\cite{alidoust2021} From the technical and fundamental physics perspectives, the MMTB approach discussed in this paper has a very low computational cost, it provides an explicit Hamiltonian that can accommodate an external magnetic field, many-body interactions, impurities and disorder, and thereby, it facilitates real space simulations, which are relevant for experiments. Furthermore, this approach provides highly deep and fundamental physical insights into the electronics properties and interactions that can occur in a system such as the detailed orbital interactions briefly discussed and shall be presented at length in the following.               

The paper is organized as follows. In Sec.~\ref{config}, we discuss the configuration, crystal symmetry of $\rm Mg_2X$, and the displacement vectors used in our MMTB models. In Sec.~\ref{lowEHamil}, the TB formalism is summarized. In Sec.~\ref{MC} and \ref{DFT}, the Monte-Carlo search algorithm and details of the DFT calculations are given, respectively. The results are presented in Sec.~\ref{results} and finally, the concluding remarks are summarized in Sec.~\ref{conclusions}.

\begin{figure}[t!]
\includegraphics[width=0.45\textwidth]{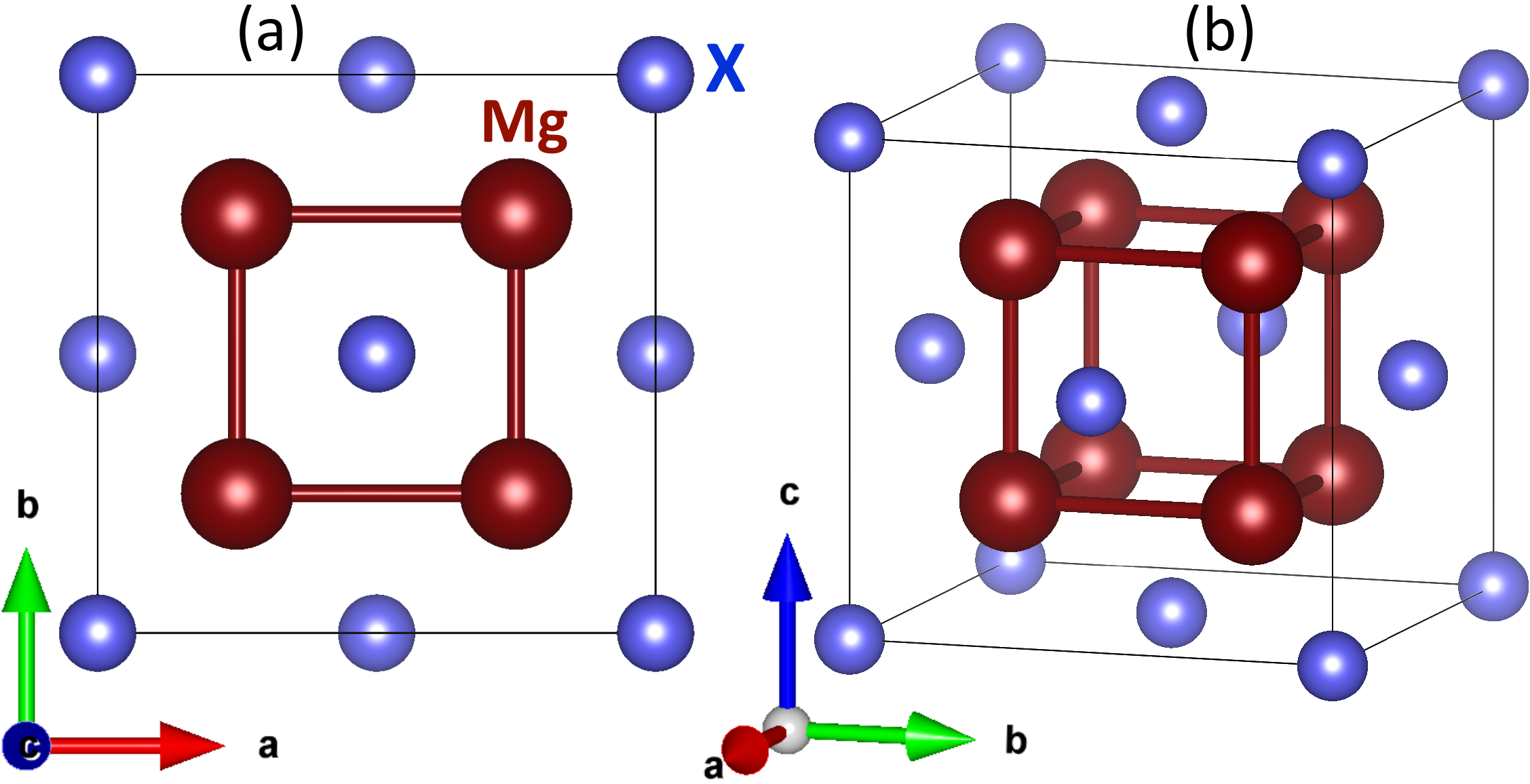}
\caption{\label{diag} The configuration of the $\rm M_2X$ compounds. The magenta and blue spheres display the $\rm Mg$ and $\rm X$ atoms, respectively. Panel (a) shows the top view of the structure and panel (b) is a three-dimensional view of the same structure. The principal axes are marked by $\rm a,b,c$. The lattice constant is $a$, which connects two $\rm X$ atoms along each principal axis.}
\end{figure}

\section{Configuration of $\rm Mg_2X$}\label{config}
Figure \ref{diag} displays the configuration of $\rm Mg_2X$ with two different views. Figure \ref{diag}(a) shows the top view of the structure whereas Fig.~\ref{diag}(b) is a three-dimensional perspective view. The structure is equivalent to the fluorite structure with the anions and cations swapped and is known as antifluorite. An antifluorite configuration can be generated with a face-centered cubic Bravais lattice and a simple-cubic lattice in the interstitial sites. The $\rm X$ atoms are located on the atom sites of the former lattice while the $\rm Mg$ atoms occupy the atom sites of the latter lattice. In what follows, for simplifying our notation, we have labeled the cation sublattice sites by $\rm A (A')$ (two atoms per unit cell) and the anion sublattice site by $\rm B$ (one atom per unit cell). As seen in Fig.~\ref{diag}, the unit cell of $\rm Mg_2X$ contains twelve atoms (8:4 ratio).
To construct an accurate TB model, one needs to account for both the different atom types available in a composition and the distance and orientation of different atoms with respect to other atoms. Defining the tight-binding interaction vectors, we have summarized the nearest-neighbor interaction vectors $\boldsymbol{\delta}_{\alpha\beta}$ among atoms located on the different sublattices of $\rm Mg_2X$ in Table~\ref{tab_vc}. To further simplify our notation, we have defined $\alpha,\beta\in \{ {\rm A},{\rm A'},{\rm B} \}$. Although, throughout the paper, we shall discuss the model constructed and results obtained by the nearest-neighbor interactions, we have constructed models incorporating both the nearest-neighbor and next nearest-neighbor interactions simultaneously. Our results revealed only a slight improvement and therefore we avoid presenting the models with the next nearest-neighbor interactions.    

\begin{table}[t]
        \begin{tabular}{|cc|c|cc|c|}
    \hline
    \hline
         $\alpha$ & $\beta$ & $(x , y , z)$ & $\alpha$ & $\beta$ & $(x , y , z)$  \\
    \hline
         $\rm A$ & $\rm A'$ & $a$($\pm 1$, 0, 0)/2 & $\rm B$ & $\rm B$ & $\pm$$a$(0, 1, 1)/{2} \\
         $\rm A'$ & $\rm A$ & $a$(0, $\pm 1$, 0)/2 &  &  &$\pm$$a$(1, 0, 1)/{2} \\
          &  & $a$(0, 0, $\pm 1$)/2 &  & &  $\pm$$a$(1, 1, 0)/{2} \\
          &  &  &  &  & $\pm$$a$(0, 1, -1)/{2} \\
          &  &  &  &  &  $\pm$$a$(1, 0, -1)/{2} \\
          &  &  &  &  &  $\pm$$a$(1, -1, 0)/{2} \\
          \hline
         $\rm A$ & $\rm B$ & $a$(1, -1, -1)/4 & $\rm A'$ & $\rm B$ & $a$(-1, 1, 1)/4 \\
         $\rm B$ & $\rm A'$  & $a$(-1, -1, 1)/4 & $\rm B$  & $\rm A$   &$a$(1, 1, -1)/4 \\
           &   & $a$(-1, 1, -1)/4 &  &   & $a$(1, -1, 1)/4 \\
           &   & $a$(1, 1, 1)/4 &  &   & $a$(-1, -1, -1)/4 \\
        \hline
        \hline
        \end{tabular}
        \caption{ Nearest-neighbor vectors $\boldsymbol{\delta}_{\alpha \beta}$ among the sites of the sublattices $\rm A$, $\rm A'$, $\rm B$. The lattice constant is denoted by $a$.
    } 
\label{tab_vc}
\end{table}

\section{Multi-orbital tight-binding model}\label{lowEHamil}
When isolated atoms are brought together to form a material, the individual atomic orbitals of each atom start interacting with those of neighboring atoms. This interaction is largest among electrons that occupy the outer electron shells and diminishes among electrons within inner electron shells. The interactions can be bonding or anti-bonding and therefore, the molecular orbitals that describe the physical properties of a material can be a combination of the original atomic orbitals. To construct a reliable model Hamiltonian that contains these pivotal aspects, the interaction among the electrons, occupying the outer electron shells, should be properly accounted for. In this case, the Hamiltonian can be expressed by   
\begin{align}\label{eq:Ham}
    \hat{\mathcal{H}}=\sum_{\mathbf{k},\mu\nu,\alpha\beta,\sigma\sigma'}&\Big\{ \mathcal{E}^{\alpha\beta}_{\mu\nu,\sigma\sigma'}\delta_{\alpha\beta} \delta_{\mu\nu}\delta_{\sigma\sigma'}\nonumber +\gamma_{\mu\nu,\sigma\sigma'}^{\alpha\beta}(\mathbf{k})\delta_{\sigma\sigma'} \\ &+\eta_{\mu\nu,\sigma\sigma'}^{\alpha\beta}\delta_{\alpha\beta} \Big\}{C_{\mu\nu,\sigma\sigma'}^{\alpha\beta\dagger}} C_{\mu\nu,\sigma\sigma'}^{\alpha\beta} + \text{h.c.},
\end{align}
where the on-site energy, hopping integrals among atoms, and spin-orbit interaction are denoted by $\mathcal{E}$, $\gamma$, and $\eta$, respectively. The different atoms are marked by $\alpha,\beta$ whereas $\mu,\nu$ and $\sigma,\sigma'$ represent the electron orbitals and spin species, respectively, ${C_{\mu\nu,\sigma\sigma'}^{\alpha\beta(\dagger)}} $ is the annihilation (creation) operator and $\delta_{ij}$ is the Kronecker delta function. By the inclusion of multi-orbital interactions among different atoms, electrons can occupy mixed and intermediate states with some finite probability. The multi-orbital nature of our tight-binding Hamiltonian is encoded into the hopping integrals as follows
\begin{equation}\label{eq:gam}
    \gamma^{\alpha\beta}_{\mu\nu}(\mathbf{k})\equiv \sum_{\boldsymbol{\delta} \in \boldsymbol{\delta}_{\alpha \beta}}t^{\alpha\beta}_{\mu\nu}(\boldsymbol{\delta})\exp({i\mathbf{k}\cdot \boldsymbol{\delta}}),
\end{equation}
in which $\mathbf{k}$ is the momentum of a particular moving electron, $\boldsymbol{\delta}$ is the displacement vector, and $t^{\alpha\beta}_{\mu\nu}(\boldsymbol{\delta})$ is the hopping integrals \cite{SK}. 

Because of stronger interactions in heavy elements, the spin-orbital mediated interactions play an important role and considerably change the properties of a material.\cite{winkler} The intrinsic spin-orbit interaction in a crystal can generally be given by\cite{winkler}
\begin{subequations}
\begin{eqnarray}\label{eq:Hso}
   \mathcal{H}_\text{SO}=\zeta(r)\hat{\mathbf{L}}\cdot\hat{\mathbf{S}}, \\
    \zeta(r)\propto \frac{1}{r}\frac{\partial V(r)}{\partial r},
   \end{eqnarray}
   \end{subequations}
where $\mathbf{L}$ and $\mathbf{S}$ are the total orbital and spin angular momentum operators, respectively, and $V(r)$ is the total crystal potential. In this model, $\zeta(r)$ depends both on position $r$ and the crystal potential, and therefore deals with the radial part $R_{n,l}(r)$ of the electron-orbital wave functions $\Psi_{n,l,m}$, which are dependent on quantum numbers $n$ and $l$ through
\begin{equation}\label{eq:wf}
    \Psi_{n,l,m}=R_{n,l}(r)Y_{l,m}(\theta ,\phi).
\end{equation}  
 Therefore, the spin-orbit coupled part of the Hamiltonian Eq.~(\ref{eq:Ham}) in its component form can be expressed by 
 \begin{equation}\label{eq:HSO}
    \eta_{\mu\nu,\sigma\sigma'}=\eta_\text{SO}\left\langle \hat{\mathbf{L}}\cdot\hat{\mathbf{S}}\right\rangle_{\mu\nu,\sigma\sigma'}.
\end{equation}
To evaluate Eq.~(\ref{eq:HSO}), it is more convenient to introduce the ladder SOI operators as follows
\begin{equation}\label{eq:lad}
\begin{gathered}
     \hat{\mathbf{L}}\cdot\hat{\mathbf{S}}= \hat{\text{L}}_z\hat{\text{S}}_z+\frac{1}{2}(\hat{\text{L}}_+\hat{\text{S}}_- + \hat{\text{L}}_-\hat{\text{S}}_+),\\
\hat{\text{L}}_\pm =\hat{\text{L}}_x\pm i \hat{\text{L}}_y,\\
\hat{\text{S}}_\pm =\hat{\text{S}}_x\pm i \hat{\text{S}}_y.
\end{gathered}    
\end{equation}
Defining $n,l,m,s$ as the principal, azimuthal, magnetic, and spin quantum numbers, respectively, the quantum numbers are restricted by $l=\{0,1,2,..,n-1\}$, $m=\{-l,-l+1,...,l-1,l\}$, and $s=\pm 1/2$. 
The operation of the angular and spin operators on a wave function at a state $\psi_{l,m,s}$ are summarized as follows
\begin{equation}\label{eq:ladOps}
    \begin{gathered}
        \hat{\text{S}}_z \psi_{l,m,s}= s\psi_{l,m,s},\\
          \hat{\text{L}}_z \psi_{l,m,s}= m\psi_{l,m,s}, \\
        \hat{\text{L}}_\pm \psi_{l,m,s}= \sqrt{(l\mp m)(l\pm m+1)}\psi_{l,m\pm 1,s}, \\
        \hat{\text{S}}_\pm \psi_{l,m,\pm1/2}=0, \\
         \hat{\text{S}}_\pm \psi_{l,m,\mp1/2}=\psi_{l,m,\pm1/2},\\
         \hat{\text{L}}_z\hat{\text{S}}_z\psi_{l,m,s} = ms\psi_{l,m,s},\\
         \hat{\text{L}}_\pm\hat{\text{S}}_\mp\psi_{l,m,\pm 1/2} = \frac{1}{2}\sqrt{(l-m)(l+m+1)}\psi_{l,m\pm 1,\mp 1/2}.
    \end{gathered}
\end{equation}
Further details on the inclusion of SOI in the formalism is presented Appendix~\ref{apx:hop} (see Eq. (\ref{eq:gamma_SOC})). In the equations above, we have set $\hbar=1$ to simplify the notation.

Since the inner shell electrons are strongly bonded to nuclei, one can consider them as frozen electrons and only account for the interaction among the valence shell electrons. Therefore, we have constructed several models, accounting for excited electrons up into the $d$ orbitals, i.e., $\{d_{xy},d_{yz},d_{zx}, d_{x^2-y^y},d_{3z^2-r^2}\}$. For example, in the smallest model where it describes five electronic bands around the Fermi level, we consider the interaction among $\{s,p_x,p_y,p_z\}$ orbitals. In the largest model, the interaction among $\{s,p_x,p_y,p_z,d_{xy},d_{yz},d_{zx}, d_{x^2-y^y},d_{3z^2-r^2},s^*\}$ orbitals are considered and the model is able to describe eighteen electronic bands around the Fermi level. These interactions result in relatively large expressions for the hopping integrals that are given in Appendix~\ref{apx:hop}.  

\section{Machine-learning for finding optimized parameters to the TB models}\label{MC}

As described in the previous section and given in Appendix \ref{apx:hop}, the TB models in the presence of SOI can contain tens of unknown parameters to be calibrated through comparison with a reference data set (in our case, the DFT data of the band structure is the reference). Therefore, one needs an efficient approach to find optimal and reliable parameter values to the TB models. One efficient machine-learning approach is the simulated annealing (SA), which is a Monte-Carlo method for derivative-free optimization, built on concepts from statistical physics \cite{AIintroBok}. The SA algorithm mimics the annealing process by first defining some cost function. Next, an initially high temperature $T$ is simulated by allowing the coordinates of the system to randomly fluctuate. Thermal equilibrium is gradually achieved by slowly cooling the system and primarily allowing fluctuations that decrease the cost-function. Fluctuations that increase the cost-function are accepted with probability $h(\Delta \mathbb{C})$. In our optimization process, we have defined a cost function
\begin{equation}\label{eq:RMSD}
\begin{gathered}
    \mathbb{C}(\mathcal{E})=\sqrt{\frac{1}{M}\sum_{i}w_{i}\left(\mathcal{E}_{i}-\mathcal{E}^\text{r}_{i}\right)^2}, \\
     M=\sum_{i}w_{i}.
     \end{gathered}
\end{equation}
In Eq.~(\ref{eq:RMSD}), the sum runs both over all the data-points of the model and the corresponding reference data points $\mathcal{E}_i^\text{r}$. Depending on the problem, the weight distribution can be set to $w_i=1$. Oftentimes, a nonuniform weight distribution, such as increased weights for bands closer to the Fermi energy, is appropriate for problems where the deviation among the data points and reference data points is considerably large and produces outliers. 
The acceptance criterion of a set of parameter values is used as follows 
\begin{equation}\label{eq:accept}
    h(\Delta \mathbb{C})=\frac{1}{1+\text{exp}(\Delta \mathbb{C}/T)}\approx \text{exp}(-\Delta \mathbb{C}/T).
\end{equation}
The variation of $\mathbb{C}$ between two steps is shown by $\Delta \mathbb{C}$. The cooling schedule for a given parameter $\alpha^i$ is 
\begin{equation}\label{eq:VFA cooling}
    T(t)=T_{0}\exp\left(-c t^{1/d}\right),
\end{equation}
where $T_0$ is the initial temperature, $d$ is the (effective) dimension of parameter space, and $c$ is a tunable cooling constant. 
To incorporate the temperature schedule, i.e., Eq.~(\ref{eq:VFA cooling}), the update to each parameter $\alpha^i$
\begin{equation}\label{eq:param update}
    \alpha_{t+1}^{i}=\alpha_{t}^{i}+y^i(\mathbb{B}_i-\mathbb{A}_i),
\end{equation}
are drawn from the distribution 
\begin{equation}\label{eq:ASA distr}
\begin{gathered}
    y^i =\text{sgn}\left(u^i-\frac{1}{2}\right)T(t_i)[(1+T(t_i)^{-1})^{|2u^i-1|}-1] ,\\
    u^{i}=U[0,1].
    \end{gathered}
\end{equation}
where $\rm sgn$ is the sign function, $T(t_i)$ is the temperature of parameter $\alpha^i$ at step $i$, and $U[0,1]$ is the continuous uniform distribution between 0 and 1. Also, $\mathbb{B}_i$ and $\mathbb{A}_i$ are the upper and lower boundaries for the search space of parameter $\alpha^i$. The random generator (\ref{eq:ASA distr}) always produces a number in the range $[-1,1]$. 
The full span of the search space for each parameter $\mathbb{B}_i-\mathbb{A}_i$ is not necessarily known, but may be approximated to bias the algorithm towards generating step sizes proportional to the relevant search space. 
Fluctuations in more sensitive parameters cause more significant changes to the cost function. One can efficiently optimize values to the more sensitive and the less sensitive parameters by multiple rounds of the annealing process, where more (less) sensitive parameters are gradually allowed to fluctuate relatively less (more). This re-annealing strategy was suggested by Ingber \cite{SAcompare} where the less sensitive parameters are periodically re-annealed as follows
\begin{equation}\label{eq:reanneal time}
    \begin{gathered}
        t\leftarrow t_i'=\text{max}[0,\Big(\frac{1}{c}\ln(\frac{T_0}{T(t_i)}\frac{s_i}{s_{\text{max}}})\Big)^d ], \\
        s_i=(\mathbb{B}_i-\mathbb{A}_i)\frac{\partial \mathbb{C}(\boldsymbol{\alpha})}{\partial \alpha^i}\\
        \approx (\mathbb{B}_i-\mathbb{A}_i) \frac{\mathbb{C}([\alpha^1,\dots,\alpha^i+\delta\alpha,\dots,\alpha^d])-\mathbb{C}(\boldsymbol{\alpha})}{\delta\alpha}.
    \end{gathered}
\end{equation}
Here, $s_i$ is the sensitivity of parameter $\alpha^i$, $s_{\text{max}}$ is the largest sensitivity, $\delta\alpha$ is a small increment in one parameter and $\boldsymbol{\alpha}$ is the vector of parameters. 

\begin{figure}[t!]
\includegraphics[width=0.35\textwidth]{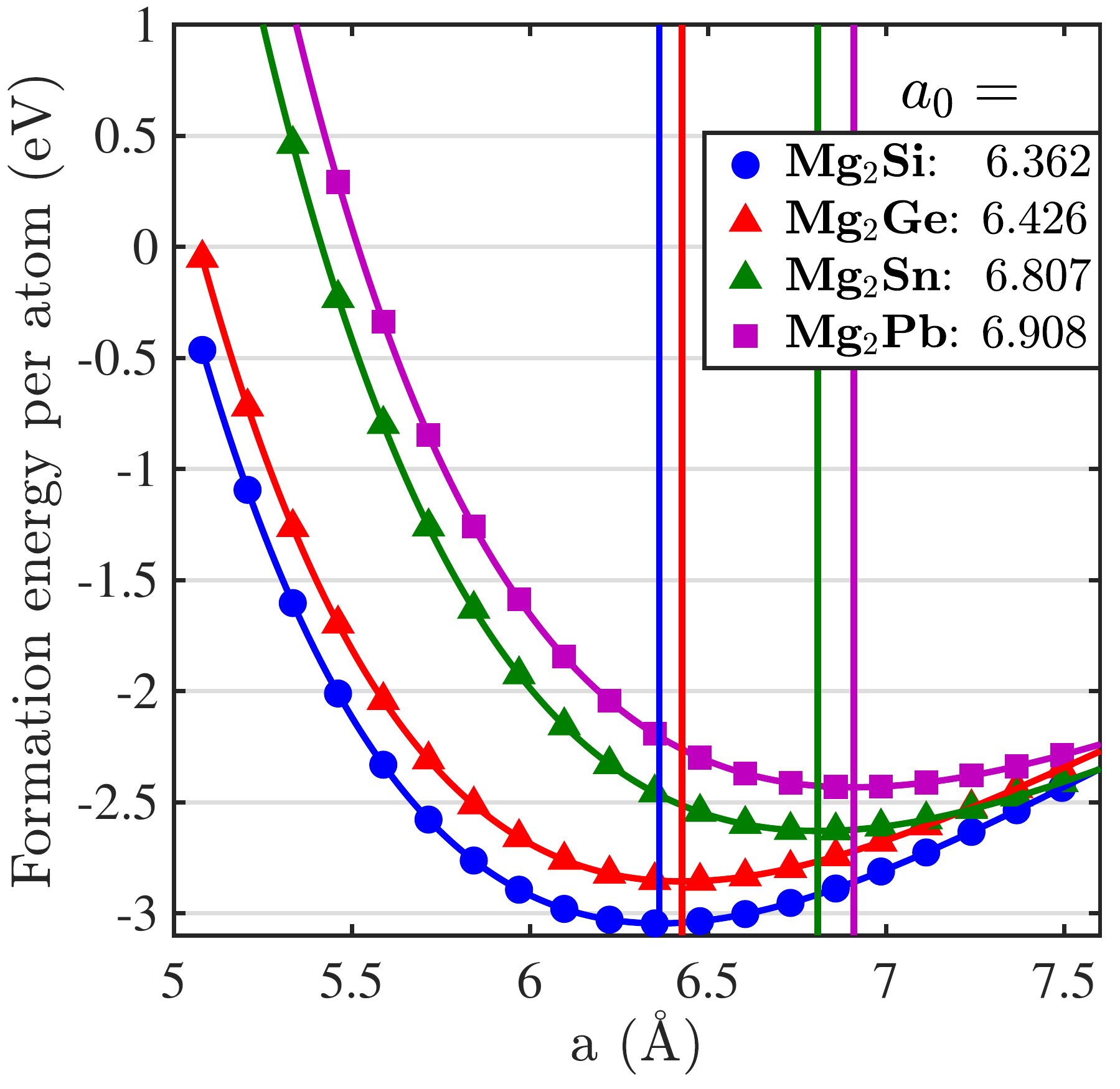}
\caption{\label{lattcons} The DFT formation energy per atom as a function of lattice parameter $a$ for the different compounds of $\rm Mg_2X~(X=Si,Ge,Sn,Pb)$. The energetically favorable lattice constants are obtained as $a_0=6.362, 6.426, 6.807, 6.908$~\AA, respectively. The vertical lines exhibit the location of the optimized lattice constants.}
\end{figure}

The required extensive search of the high-dimensional parameter space of our TB models using the Monte-Carlo algorithm is a slow process, especially when many different datasets are to be considered. To speed up our extensive search to find best optimized TB parameters, we have implemented a massively parallelized algorithm using graphical processing units (GPU).  

\section{First-principles density functional theory calculations}\label{DFT}

The density functional theory (DFT) calculations of electronic structure are performed using the $\rm GPAW$ package. \cite{gp1} We have employed the gradient-corrected PBE approximation for the exchange-correlation energy functional. The ${\bf k}$ space grid is constructed by the Monkhorst-Pack scheme and a relatively large value of grid density is used, i.e., $8.0$~${\bf k}$-points per $1/\text{\AA}$. The kinetic energy cutoff for plane waves is set to $800$~eV. The width of the Fermi-Dirac distribution is set to $0.01$~eV. All DFT calculations presented are performed in the presence of SOI. Also, the results from DFT without SOI are used for investigating the influence of SOI on the band-structure. 

To simulate the strained configurations, we introduce a strain tensor with components $\varepsilon_{ij}$ in which the indices run over real space coordinations, i.e., $ij\in {x,y,z}$. The strained unit cell and therefore updated locations of the different atoms can be described by the new vectors  
 $a=\varepsilon_{xx}  a_0$,
$b = \varepsilon_{yy}  b_0$, and
$c =\varepsilon_{zz} c_0$.
The unstrained vectors are $a_0$, $b_0$, and $c_0$. In what follows, we restrict our simulations to a low-strain regime, i.e., $\pm 10\%$ to make sure no structural transition occurs upon inserting strain into the unit cell. In order to determine any structural transition, one has to perform high-throughput calculations and span a huge search space made of various structures. Nevertheless, from previous studies on two-dimensional materials (see Ref.~\onlinecite{M.Alidoust2} and references therein), which are more sensitive to strain, we assume that the antifluorite $\rm Mg_2X$ compounds with a cubic-like configuration are stable in the presence of strain less than $\pm 10\%$. Note that $\pm 10\%$ of strain corresponds to $10\%$ of tensile and compressive strain, respectively. Throughout the paper, we consider uniform strain in all directions.

\begin{table}
\begin{tabular}{ |c||cc| } 
 \hline
  \hline
 Compound & Experiment & DFT \\ 
  & $a_0$(\AA) & $a_0$(\AA) \\
  \hline
 $\rm Mg_2Si$ & 6.340 & 6.362 \\ 
 $\rm Mg_2Ge$ & 6.385 & 6.426 \\
 $\rm Mg_2Sn$ & 6.765 & 6.807 \\
 $\rm Mg_2Pb$ & 6.836 & 6.908 \\
 \hline
  \hline
\end{tabular}
        \caption{ The lattice constant of $\rm Mg_2X$ from experiment and DFT predictions. } \label{table_lattcons}
\end{table}

The radii of $\rm X$ elements differ considerably. Therefore, one can expect that the lattice constants of $\rm Mg_2X$ compositions vary significantly. In order to find the energetically stable lattice constant for each composition, we have performed DFT calculations and plotted the formation energy of these compounds as a function of lattice constant $a$ in Fig.~\ref{lattcons}. The formation energy is normalized by the number of atoms in the unit cell. As seen, $\rm Mg_2Si$ has the smallest lattice constant, i.e., $a_0=6.362$\AA~ whereas the lattice constant of $\rm Mg_2Pb$ is largest: $a_0=6.908$\AA. The predicted lattice constants for the compounds $\rm Mg_2Ge$ and $\rm Mg_2Sn$ are $a_0=6.426, 6.807$\AA, respectively. These results are perfectly aligned with the increase of atomic numbers and radius of atoms among $\rm Si,Ge,Sn,Pb$. The obtained lattice constants are in good agreement with experiments as summarized in Table~\ref{table_lattcons}. \cite{U.Winkler,so1_exp,so2_exp} In the calculations and discussions that follow, we shall use the above predicted lattice constants for the relaxed structures in the absence of strain and obtain strained structures accordingly. Therefore, the normalized formation energy of the strained structures can be found in Fig.~\ref{lattcons} as well.

\begin{figure*}[t!]
\includegraphics[width=\textwidth]{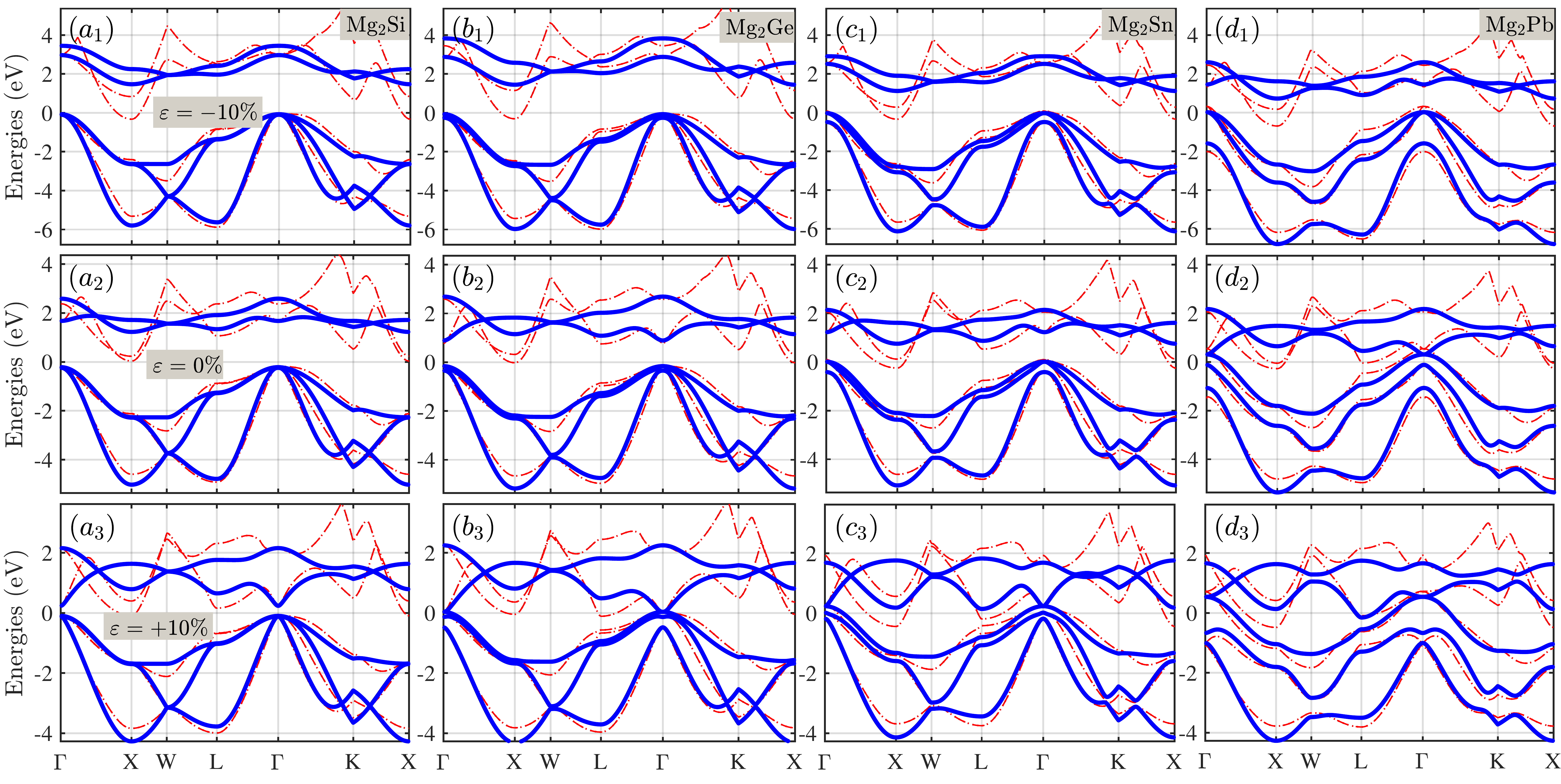}
\caption{\label{bs_5x} The band structure of $\rm Mg_2X$ along a high-symmetry path $\rm \Gamma XWL\Gamma KX$. Column-wise, from left to right, the element $\rm X=Si,Ge,Sn,Pb$ changes, respectively. Row-wise, the applied strain varies $\varepsilon=-10\%, 0\%, +10\%$ from top to bottom, respectively. The dashed curves show DFT predictions and the solid curves are the results of the machine-learned five-band TB model. }
\end{figure*}

\section{results and discussions}\label{results}

In Sec.~\ref{fiveTB} we discuss the band structure results of the five-band TB model. The contribution of the $s$ and $p$ orbitals of $\rm Mg$ and $\rm X$ to the total band structure is discussed in Sec.~\ref{sp_cont}. The strength of SOI and the $\sigma$ and $\pi$ bondings in $\rm Mg_2X$ are presented in Secs.~\ref{soi} and \ref{sigpi}, respectively. 

\subsection{The band structure from the machine-learned five-band TB model}\label{fiveTB}
Figure~\ref{bs_5x} exhibits the band structure of $\rm Mg_2X~(X=Si,Ge,Sn,Pb)$. In all cases, the Fermi level is shifted to zero energy. The dashed red curves are obtained through DFT whereas the solid blue curves are the results of our machine-learned five-band TB model. Column-wise, the element $\rm X$ changes as $\rm X=Si,Ge,Sn,Pb$ from left to right, respectively. Row-wise, Fig.~\ref{bs_5x} shows how strain affects the band structure at $\varepsilon=-10\%, 0\%, +10\%$ from top to bottom, respectively. The band structure is plotted along the high-symmetry path $\rm \Gamma XWL\Gamma KX$ in ${\bf k}$-space. As seen, our five-band TB model reproduces the band structures in a good agreement to the DFT results. Specifically, the three valence bands produced by the five-band TB model deviate only slightly from those of DFT. Similarly, the same precision in predicting the valence bands is accessible to larger models than the five-band model. The two conduction bands, however, have larger deviations from those of DFT although at the $\Gamma$ point we yet see good agreement between the model prediction and DFT. The main reason for the larger deviation of the conduction bands originates from the exclusion of higher excited states in our five-band TB model. In fact, an excited mode over the Fermi level is a complicated hybridization of several excited states. However, in the five-band TB model, we have considered the contribution of five orbitals only, namely the $3s$ orbital of $\rm Mg$ and the $\{p_x,p_y,p_z \}$ orbitals of the $\rm X$ elements. Therefore, in order to obtain more accurate predictions for the conduction bands using the TB method, one needs to take more of the excited orbitals into account. To confirm this, we have considered the $\{3s,3p\}$ orbitals of $\rm Mg$ and $\{s,p,d,s^* \}$ orbitals of X, constructed an eighteen-band TB model, and calibrated the model to the DFT band structures. The results revealed almost perfect reproduction of the band structures obtained by DFT, close to the Fermi level. A representative band structure from the machine-learned eighteen-band TB model is shown in Fig.~\ref{bs_18x} of Appendix \ref{apx:bs}. As seen, most of the band features and details of DFT are now reproduced with the larger TB model. Therefore, the five-band model supports the valence bands accurately and, compared to the single-band parabolic model, it captures the curvature of the valence/conduction bands at the $\gamma$ point. Depending on the required accuracy within the conduction bands, one can employ either of the five-band or eighteen-band models at the cost of computational speed. In the following, the $s$ orbital of $\rm Mg$ and the $p$ orbitals of $\rm X$ is equivalent to the $3s$ orbital of $\rm Mg$, and $3p$ orbitals of $\rm Si$, $4p$ orbitals of $\rm Ge$, $5p$ orbitals of $\rm Sn$, and $6p$ orbitals of $\rm Pb$, respectively. 

\begin{figure*}[t!]
\includegraphics[width=\textwidth]{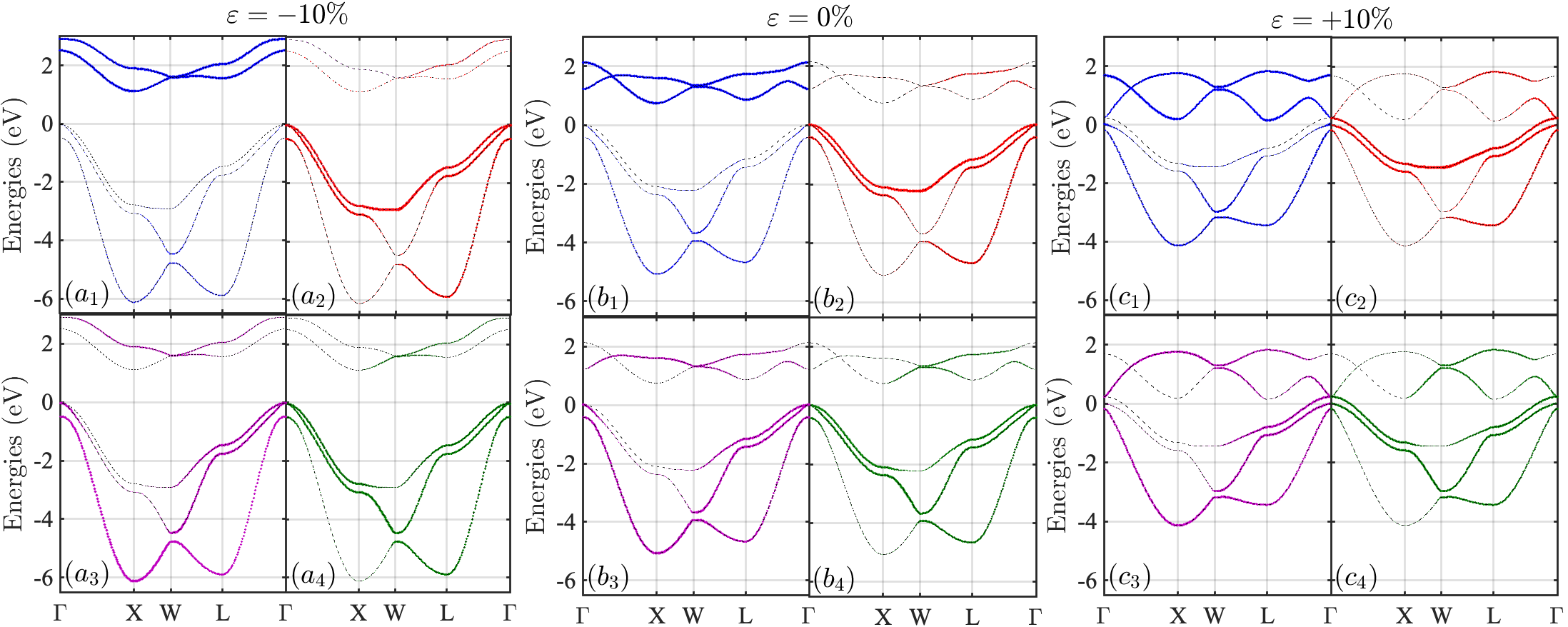}
\caption{\label{contrb_5x} The projection of the total band structure onto the $\{ s, p_x,p_y,p_z\}$ orbitals. The applied uniform strain to the system is $\varepsilon=-10\%,0\%,+10\%$ in panels with (a), (b), and (c) labels, respectively. The panels with index `1' show the contribution of the $3s$ orbital of $\rm Mg$ whereas the panels with indices `2', `3', and `4' display the contribution of the $5p_x,5p_y,5p_z$ orbitals of $\rm Sn$, respectively. }
\end{figure*}

\subsection{The contribution of $s$ and $p$ orbitals}\label{sp_cont}

The obtained parameter values of the five-band TB model for $\varepsilon=0\%,\pm 10\%$ are given in Appendix~\ref{apx:cs}. Our five-band TB model is able to describe the low-energy properties of the compounds $\rm Mg_2X$, e.g., through projection of the total band structure onto different orbitals. Figure~\ref{contrb_5x} illustrates the projection of the total band structure of $\rm Mg_2Ge$ onto each orbital. The analysis of the $\rm Mg_2X$ compounds showed that these orbitals play similar roles in them. Therefore, we discuss the representative case of $\rm Mg_2Ge$. Figures~\ref{contrb_5x}(a), \ref{contrb_5x}(b), and \ref{contrb_5x}(c) illustrate how strains of $\varepsilon=-10\%, 0\%, +10\%$, respectively, influence the contribution of the individual orbitals to the total band structure. For visualization reasons, the contribution of the bands is proportional to their thickness, i.e., the thicker parts have larger contribution than the thinner segments. Figures~\ref{contrb_5x}($\rm a_1$), \ref{contrb_5x}($\rm b_1$), and \ref{contrb_5x}($\rm c_1$) display the contribution of the $s$ orbital of $\rm Mg$ whereas the contributions of the $\{p_x,p_y,p_z\}$ orbitals of $\rm Sn$ are shown in Figs.~\ref{contrb_5x}($\rm a_2$,$\rm b_2$,$\rm c_2$), \ref{contrb_5x}($\rm a_3$,$\rm b_3$,$\rm c_3$), and \ref{contrb_5x}($\rm a_4$,$\rm b_4$,$\rm c_4$), respectively. 
\\*
First, the figures illustrate that the contribution of the $s$ orbital of $\rm Mg$ is largest to the conduction bands whereas the $p$ orbitals of $\rm X$ have the largest contribution to the valence bands. Therefore, in a compound $\rm Mg_2X$, the energy of particles occupying the $s$ orbital of $\rm Mg$ is larger than those occupying the $p$ orbitals of $\rm X$. The analysis of twelve-band and eighteen-band TB models showed that the energy of particles occupying the $s$ orbital of $\rm X$ is lower than its $p$ orbitals and thus has no contribution to the orbitals of our five-band TB model. It was found that the $p$ orbitals of $\rm Mg$ and $d$ orbitals of $\rm X$ contribute slightly to the total valence band as well, but much less significant than the $s$ orbital of $\rm Mg$ and $p$ orbitals of $\rm X$. It is well understood that the excited particles are basically governed by the conduction bands. Hence, Fig.~\ref{contrb_5x} reveals the fact that when the two elements $\rm Mg$ and $\rm X$ are brought together in the antifluorite configuration shown in Fig.~\ref{diag}, $\rm Mg$ tends to transfer its two $s$ electrons to the $p$ orbitals of $\rm X$. Therefore, in the $\rm Mg_2X (X=Si,Ge,Sn,Pb)$ compounds considered here, $\rm Mg$ plays a cation role whereas $\rm X$ turns into an anion and create $\rm Mg^{2+}_2X^{4-}$, which is consistent with the electronegativity scale.
\\*   
Second, Fig.~\ref{bs_5x} shows that the main contribution to the valence band structure and accordingly, the main physical properties of $\rm Mg_2X$ at equilibrium originate from the $p$ orbitals of the $\rm X$ atoms. Thus, to account for SOI in our five-band TB model, one can simply include the spin-orbit mediated interactions among the $p$ particles only.  

\begin{figure}[b!]
\includegraphics[width=0.40\textwidth]{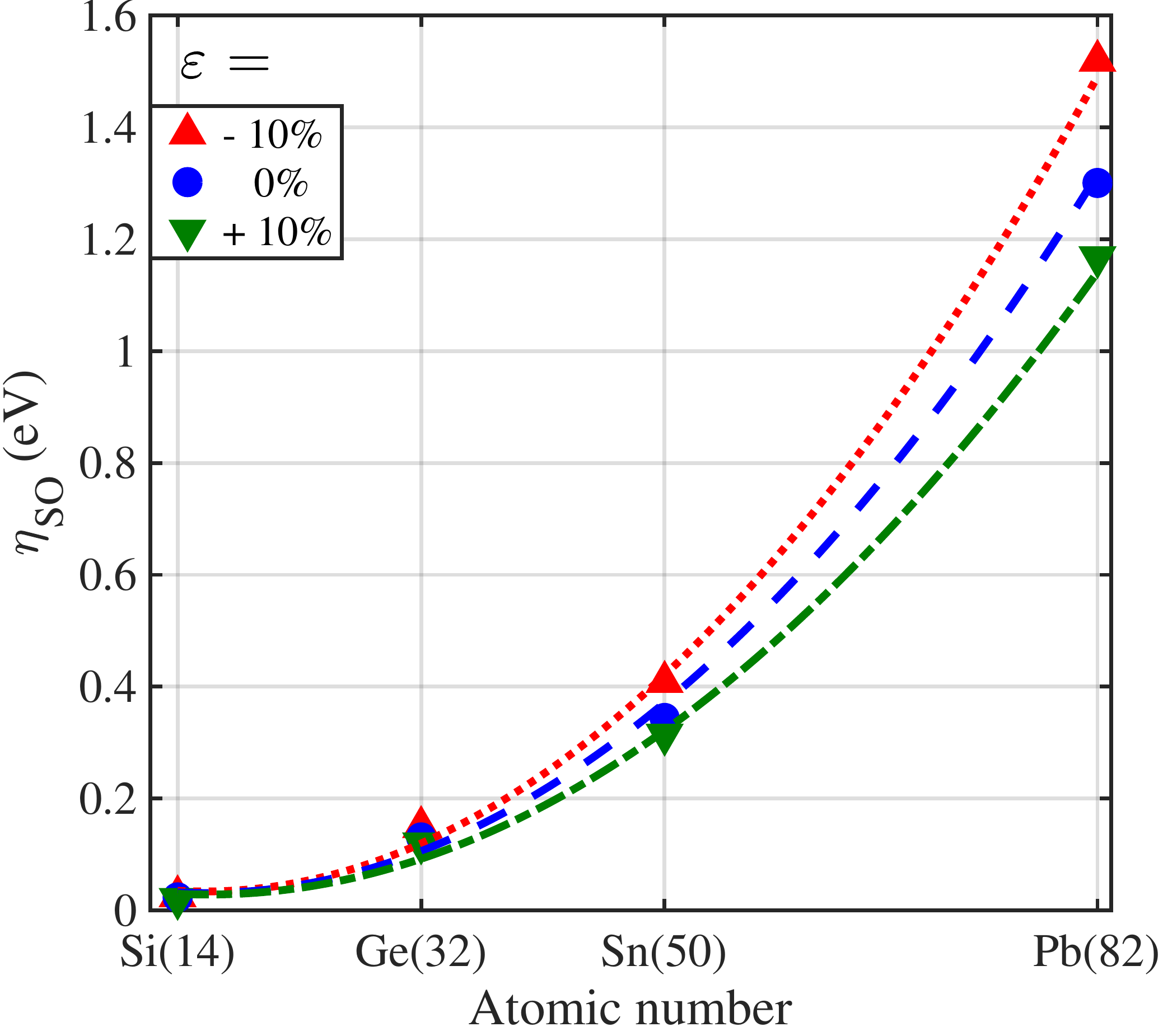}
\caption{\label{esoc} The strength of spin-orbit interaction as a function of the atomic number of the elements $\rm X=Si,Ge,Sn,Pb$. The maximum compressive and tensile strain values are applied to the system, i.e., $\varepsilon = \mp 10\%$ and compared with the cases without any strain, i.e., $\varepsilon = 0\%$.}
\end{figure}
\begin{figure*}[bth!]
\includegraphics[width=\textwidth]{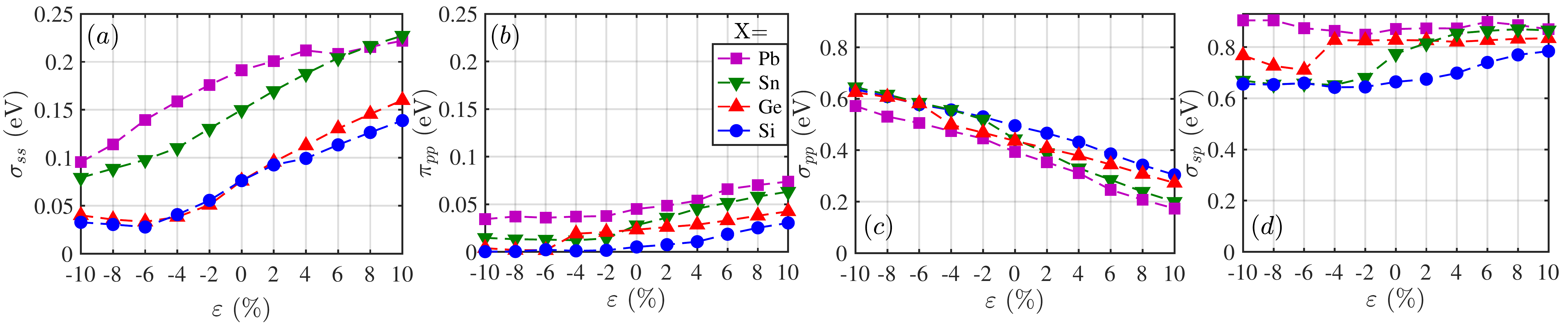}
\caption{\label{sigpi} The absolute values of bondings $\sigma_{ss}, \pi_{pp}, \sigma_{pp},  \sigma_{sp}$, as a function of strain $\varepsilon$, are displayed in panels (a)-(d), respectively. The curves with different symbols show the different compounds $\rm Mg_2X (X=Si,Ge,Sn,Pb)$.}
\end{figure*}

A compressive strain on the order of $\varepsilon=-10\%$, shown in the leftmost panels of Fig.~\ref{contrb_5x}, weakens the contribution of the $s$ orbital of $\rm Mg$ (Fig.~\ref{contrb_5x}($\rm a_1$)) and causes a dominant contribution of the $p$ orbitals of $\rm X$ to the valence bands. Subject to $\varepsilon=-10\%$ of strain, the contribution of the $p$ orbitals of $\rm X$ to the conduction bands is negligible while the $s$ orbital of $\rm Mg$ governs the conduction bands. The application of $\varepsilon=+10\%$ tensile strain, however, has an exact reversed effect and causes more contribution of the $s$ orbital of $\rm Mg$ and $p$ orbitals of $\rm X$ to the valence and conduction bands, respectively. These features can be fully understood by noting the fact that compressive and tensile strains result in stronger and weaker coupling of the orbitals, respectively. Therefore, in the presence of $\varepsilon=-10\%$ strain, the $s$ orbital of $\rm Mg$ acquires stronger contribution to the conduction bands and becomes energetically less favorable (appears at larger positive energies when comparing the conduction bands of Fig.~\ref{contrb_5x}($\rm a_1$) to Fig.~\ref{contrb_5x}($\rm b_1$)). Whereas the contribution of the $p$ orbitals of $\rm X$ to the total band structure is more localized on the valence bands. This can be confirmed when comparing Figs.~\ref{contrb_5x}($\rm a_2,a_3,a_4$) with Figs.~\ref{contrb_5x}($\rm b_2,b_3,b_4$)) where the $p$ orbitals of $\rm X$ become energetically more favorable (appear at larger negative energies). In contrast, when subject to $+10\%$ of tensile strain, the $\rm Mg$ and $\rm X$ atoms are slightly decoupled and therefore the entire system tends to acquire properties closer to isolated atomic species. This is clearly seen in Figs.~\ref{contrb_5x}(c) that not only the contribution of the $s$ orbital of $\rm Mg$ to the valence bands is strengthened but the conduction bands also acquire smaller positive energies. Likewise, the $p$ orbitals of $\rm X$ acquire smaller negative energies. The results of our five-band TB model for the orbital contributions in unstrained $\rm Mg_2X$ are in agreement with those concluded from DFT. \cite{J.J.Pulikkotil,P.M.Lee,M.Y.Au-Yang}

\subsection{Strength of spin-orbit interaction}\label{soi}

As pointed out earlier, the TB model allows us to estimate the strength of SOI in the $\rm Mg_2X$ compounds. Utilizing our machine-learned five-band TB model, we have summarized the strength of SOI, $\eta_\text{SO}$, for $\rm X=Si,Ge,Sn,Pb$ and strain values from $\varepsilon=-10\%$ to $\varepsilon=+10\%$ by a step of $2\%$ in Table~\ref{tbl_so}. To obtain the results of $\eta_\text{SO}$ in Table~\ref{tbl_so}, we have confined our model to a small interval around the $\Gamma$ point. This confinement results in a perfect fit to the DFT band structure and more accurate estimation of $\eta_\text{SO}$. Also, we have summarized the values of $\eta_\text{SO}$ by fitting the five-band TB model to the entire Brillouin zone in Appendix~\ref{apx:cs}. Nevertheless, the physical trends and conclusions made remain the same. We have shown the values of $\eta_\text{SO}$, obtained through the confined five-band TB model, as a function of $\rm X=Si,Ge,Sn,Pb$ for $\varepsilon=0\%,\pm 10\%$ by dots in Fig.~\ref{esoc}. As seen, the strength of SOI follows a significant and systematic enhancement by increasing the atomic number $z$ of $\rm X$. We tested several functions, including exponential and polynomial ones, to determine the functionality of $\eta_\text{SO}$ to $z$ and $\varepsilon$. Our results suggested a second order polynomial as a function of $z$ with a least deviation from the original data points as follows 
\begin{equation}\label{eta_so}
\eta_\text{SO}(\varepsilon,z) = {\cal A}(\varepsilon) z^2+{\cal B}(\varepsilon) z+{\cal C}(\varepsilon).
\end{equation}
Further investigations found that the coefficients of the polynomial function possess linear functionalities with respect to strain $\varepsilon$ so that the coefficients can be described by
\begin{equation}\label{coff}
\begin{gathered}
{\cal A}(\varepsilon) = {\cal A}_1\varepsilon + {\cal A}_2,\\
{\cal B}(\varepsilon) = {\cal B}_1\varepsilon + {\cal B}_2,\\
{\cal C}(\varepsilon) = {\cal C}_1\varepsilon + {\cal C}_2,\\
\end{gathered}
\end{equation}
in which the constant values are ${\cal A}_1=-0.385$\text{meV}, ${\cal A}_2=0.681$\text{meV}, ${\cal B}_1=11.7$\text{meV}, ${\cal B}_2=-2.12$\text{meV}, ${\cal C}_1=-115.5$\text{meV}, ${\cal C}_2=223.3$\text{meV}. Therefore, Eqs.~(\ref{eta_so}) and (\ref{coff}) determine the dependence of the strength of SOI to the atomic number of $\rm X$ and strain $\varepsilon$. To illustrate the efficiency of this model, we have plotted $\eta_\text{SO}(\varepsilon,z)$ as a function of $z$ and three values of strain by dashed lines in Fig.~\ref{esoc}. We see that the obtained equation (\ref{eta_so}) can properly predict the variation of SOI in $\rm Mg_2X$ subject to strain. Figure~\ref{esoc} and Table~\ref{tbl_so} demonstrate that compressive and tensile strains enhance and suppress $\eta_\text{SO}$, respectively. This is consistent  with the contribution of the $p$ orbitals of $\rm X$ to the total band structure found in Fig.~\ref{contrb_5x}. 

According to the findings of thermoelectric effects reported in the literature and discussed earlier in Introduction, the delicate splitting and aspects of the band structure highly influence the Seebeck coefficient, charge conductance, and phonon scattering. The latter quantities determine the thermoelectric power of a material. Therefore, our results of the band structure upon changing $\rm X$ and exerting strain point into the fundamentally important ingredients to be accounted for when analyzing various physical properties of $\rm Mg_2X$. Note that, in general, the unstrained $\rm Mg_2X$ compounds are indirect band gap semiconductors \cite{H.Balout} and we here instead focus on the valence band splittings. The detailed study on the indirect band gaps of the $\rm Mg_2X$ compounds subject to strain can be the focus of a future work.  

The three valence bands in the total band structure are degenerate at the $\Gamma$ point for some cases as seen in Fig.~\ref{bs_5x}. These three bands however split into two and three bands at the $\Gamma$ point as a result of SOI, depending on strain and the atomic number of $\rm X$. The results are summarized in Table~\ref{tbl_so}. The highest valence band at the $\Gamma$ point is labeled by $1^\text{st}$ and the first and second bands below the highest valence band are named $2^\text{nd}$ and $3^\text{rd}$ bands, respectively. The splitting magnitudes between $1^\text{st}$-$2^\text{nd}$ and $1^\text{st}$-$3^\text{rd}$ are denoted by $g^1_{\varepsilon,\text{SO}}$ and $g^2_{\varepsilon,\text{SO}}$, respectively. The $g^1_{\varepsilon,\text{SO}}$ band splitting at the $\Gamma$ point in $\rm Mg_2Si$ is zero throughout the entire strain interval we considered. However, $g^1_{\varepsilon,\text{SO}}$ acquires nonzero values when applying $\varepsilon>5\%$ of tensile strain to $\rm Mg_2Ge$. The splitting increases in $\rm Mg_2Sn$ and $\rm Mg_2Pb$ and appears even at compressive strains. This is in contrast to $g^2_{\varepsilon,\text{SO}}$, which is nonzero for all compounds and enhances (suppresses) in the presence of compressive (tensile) strain similar to the strength of SOI. Note that the values of the band splittings $g^2_{\varepsilon,\text{SO}}$ are different from the strength of SOI, $\eta_\text{SO}$. 

\subsection{The contribution of $\sigma$ and $\pi$ bondings}\label{sigpi}
The five-band TB model further provides insightful information into the type of bondings and their strength in the $\rm Mg_2X$ compounds. We have extracted the behavior of the $\sigma$ bondings and $\pi$ bonding energies from our model and plotted them as a function of strain for different $\rm X$ in Fig.~\ref{sigpi}. Also, the exact values of the bonding parameters are given in Appendix~\ref{apx:cs}. The $\sigma$ bondings among $s$-$s$ orbitals, $p$-$p$ orbitals, and $s$-$p$ orbitals are shown by $\sigma_{ss}$, $\sigma_{pp}$, and $\sigma_{sp}$, respectively, while the $\pi$ bonding among the $p$-$p$ orbitals is denoted by $\pi_{pp}$. The results in Fig.~\ref{sigpi}(b) reveal that the $\pi_{pp}$ bonding is negligible compared with the $\sigma$ bondings shown in Figs.~\ref{sigpi}(a), \ref{sigpi}(c), and \ref{sigpi}(d). The largest variation upon strain belongs to $\sigma_{pp}$ and compressive strain enhances $\sigma_{pp}$. Also, Figs.~\ref{sigpi}(c) and \ref{sigpi}(d) show that $\sigma_{sp}$ and $\sigma_{pp}$ are the dominating bondings in $\rm Mg_2 X$.

\onecolumngrid
\begin{center}
\begin{table}
\begin{tabular}{ |c||ccccccccccc| }
 \hline
  \hline
 \multicolumn{12}{|c|}{The strength of spin-orbit interaction from the five band TB model $\eta_\text{SO}$ (meV)} \\
 \hline
 $\varepsilon$ & -$10\%$ & -$8\%$ & -$6\%$ & -$4\%$ & -$2\%$ & $0\%$ & +$2\%$ & +$4\%$ & +$6\%$ & +$8\%$ & +$10\%$\\
 \hline
 $\rm Mg_2Si$   & 25.4 & 24.7 & 24.0 & 23.6 & 23.2 & 23.0 & 22.7 & 21.3 & 21.0 & 20.6 & 20.4 \\
 $\rm Mg_2Ge$  & 149.1 & 144.8 & 141.0 & 136.9 & 133.5 & 130.5 & 127.9 & 125.7 & 124.3 & 121.8 & 119.7 \\
 $\rm Mg_2Sn$  & 408.7 & 394.3 & 379.9 & 368.3 & 355.9 & 344.1 & 332.2 & 321.2 & 320.9 & 316.5 & 314 \\
 $\rm Mg_2Pb$  & 1519.0 & 1475.9 & 1425.5 & 1376.4 & 1335.6 & 1300.5 & 1269.2 & 1241.6 & 1216.6 & 1193.8 & 1167.5 \\
 \hline
  \multicolumn{12}{|c|}{The band splitting between $1^\text{st}$ and $2^\text{nd}$ valence bands at the $\Gamma$ point $g_{\varepsilon,\text{SO}}^1$ (meV)} \\
 \hline
 $\rm Mg_2Si$   & 0.0 & 0.0 & 0.0 & 0.0 & 0.0 & 0.0 & 0.0 & 0.0 & 0.0 & 0.0 & 0.0 \\
 $\rm Mg_2Ge$  & 0.0 & 0.0 & 0.0 & 0.0 & 0.0 & 0.0 & 0.0 & 0.0 & 16.8 & 182.6 & 179.7 \\
 $\rm Mg_2Sn$  & 0.0 & 0.0 & 0.0 & 0.0 & 0.2 & 0.3 & 0.5 & 0.9 & 1.0 & 20.1 & 265.2 \\
 $\rm Mg_2Pb$  & 0.0 & 0.0 & 0.3 & 18.1 & 316.6 & 592.0 & 855.0 & 1097.1 & 1318.7 & 1519.5 & 1699.5  \\
 \hline
   \multicolumn{12}{|c|}{The band splitting between $1^\text{st}$ and $3^\text{rd}$ valence bands at the $\Gamma$ point $g_{\varepsilon,\text{SO}}^2$ (meV)} \\
 \hline
 $\rm Mg_2Si$   & 37.5 & 36.4 & 35.4 & 34.5 & 33.6 & 32.9 & 32.2 & 31.5 & 30.9 & 30.4 & 29.9 \\
 $\rm Mg_2Ge$  & 226.2 & 219.8 & 213.6 & 207.9 & 202.6 & 197.8 & 193.4 & 189.4 & 185.8 & 276.5 & 515.4 \\
 $\rm Mg_2Sn$  & 622.8 & 600.8 & 580.5 & 561.9 & 545.1 & 529.1 & 516.7 & 504.9 & 493.8 & 482.1 & 471.6 \\
 $\rm Mg_2Pb$  & 2334.5 & 2247.0 & 2166.4 & 2093.1 & 2029.8 & 1973.3 & 1918.9 & 1871.9 & 1822.3 & 1778.2 & 1748.3  \\
 \hline
  \hline
\end{tabular}
\caption{The strength of SOI, $\eta_{\text{SO}}$, and band splittings $g^1_{\varepsilon,\text{SO}}$ and $g^2_{\varepsilon,\text{SO}}$ at the $\Gamma$ point in the $\rm Mg_2X (X=Si,Ge,Sn,Pb)$ compounds subject to strain $\varepsilon\in[-10\%,+10\%]$. }\label{tbl_so}
\end{table}
\end{center}
\twocolumngrid   
        
\section{conclusions}\label{conclusions}

We have developed machine-learned multi-orbital tight-binding (MMTB) model Hamiltonians to study the electronic properties of the $\rm Mg_2X~(X=Si,Ge,Sn,Pb)$ antifluorite structures subject to uniform strain. The MMTB models are fitted to the DFT band structure using a massively parallelized Monte-Carlo search algorithm. The investigations demonstrate that a machine-learned five-band TB model, accounting for the $s$ orbital of $\rm Mg$ and $\{p_x,p_y,p_z\}$ orbitals of $\rm X$ occupying their outer electron shells, respectively, can sufficiently describe the electronic characteristics of $\rm Mg_2X$ close to the Fermi level, in particular, the valence band and the effect of SOI. We find that the $\rm Mg$ atoms tend to transfer their electrons, occupying the $s$ orbital, to the $p$ orbitals of the $\rm X$ atoms and create $\rm Mg^{2+}_2X^{4-}$. This phenomenon is clearly seen in the projected band structure where the $s$ orbital of $\rm Mg$ is largest on the conduction bands, supporting excited states, whereas the contribution of the $p$ orbitals of $\rm X$ is largest on the valence bands with energies lower than the Fermi level. The application of compressive strain causes further localization of the contribution of the $s$ orbital of $\rm Mg$ to the conduction bands at higher energies above the Fermi level and the $p$ orbitals of $\rm X$ to the valence bands at lower energies below the Fermi level. Tensile strain, however, has a reversed effect and weakens the contributions of the $s$ orbital of $\rm Mg$ and the $p$ orbitals of $\rm X$ to the conduction bands and valence bands, respectively. 

The analysis of the projected band structures shows that the spin-orbit interaction (SOI) in $\rm Mg_2X$ originates from the $p$ orbitals of $\rm X$. Extracting the strength of SOI $\eta_\text{SO}$ through the five-band TB model, we have obtained a function for predicting $\eta_\text{SO}$ in the presence of strain and against the atomic number of $\rm X$. The five-band TB model reveals that the main bonding contributions in the $\rm Mg_2X$ compounds are $|\sigma_{sp}| > |\sigma_{pp}| > |\sigma_{ss}|$, and that the $\pi$ bonding is negligible. $\sigma_{pp}$ shows the largest variation upon applying strain.  

We have found that both $\rm X$ and strain can effectively control the band gap of $\rm Mg_2X$, turning a semiconductor into metal and vice versa, and efficiently manipulate the band splitting of the valence bands at the $\Gamma$ point. These findings point into controllable electronic properties, quantum transport, and thermoelectric effects in these materials. Our five-band TB model can be generalized and utilized to study large-scale $\rm Mg_2X$-based compounds both in molecular dynamics simulations and quantum transport studies with more accuracy compared to those of single-band parabolic models available in the literature. Moreover, the presented TB construction approach which combines DFT band structure with multi-dimensional Monte-Carlo search of parameters can be applied to a wide variety of materials.   

\acknowledgements

The DFT calculations were performed using the resources provided by UNINETT Sigma2—the National Infrastructure for High Performance Computing and Data Storage in Norway, project number: NN9497K. Financial support from the NTNU Digital Transformation program (Norway) for the project ALLDESIGN is gratefully acknowledged.

\onecolumngrid
\appendix

\section{Hopping integrals and SOI matrix elements for the eighteen-band TB model}\label{apx:hop}
In this Appendix, we present the components of the largest TB model constructed. This model includes $\{s,p_x,p_y,p_z,d_{xy},d_{yz},d_{zx}, d_{x^2-y^y},d_{3z^2-r^2},s^*\}$ orbitals and describes eighteen electronic bands around the Fermi level. To construct smaller models, similar to the five-band TB model presented in the main text, one simply needs to remove the associated interactions in the eighteen-band TB model. The basis set for the largest TB model is given by 
\begin{eqnarray}\label{eq:18xbasis}
    \begin{gathered}
        \Psi_{18} = \left(  C_{s\sigma}^{\rm A}, C_{p_x\sigma}^{\rm A}, C_{p_ys\sigma}^{\rm A}, C_{p_z\sigma}^{\rm A}, C_{s\sigma}^{\rm A'}, C_{p_x\sigma}^{\rm A'}, C_{p_ys\sigma}^{\rm A'}, C_{p_z\sigma}^{\rm A'},  C_{s\sigma}^{\rm B}, C_{p_x\sigma}^{\rm B}, C_{p_y\sigma}^{\rm B}, C_{p_z\sigma}^{\rm B},\right. \\ 
        \left. C_{d_{xy}\sigma}^{\rm B}, C_{d_{yz}\sigma}^{\rm B}, C_{d_{zx}\sigma}^{\rm B}, C_{d_{x^2-y^2}\sigma}^{\rm B}, C_{d_{3z^2-r^2}\sigma}^{\rm B}, C_{s^{*}\sigma}^{\rm B} \right)^T.
    \end{gathered}
\end{eqnarray}
The basis set used in the five-band model is instead 
\begin{eqnarray}\label{eq:5xxbasis}
    \begin{gathered}
        \Psi_5 = \left(  C_{s\sigma}^{\rm A},C_{s\sigma}^{\rm A'}, C_{p_x\sigma}^{\rm B}, C_{p_y\sigma}^{\rm B}, C_{p_z\sigma}^{\rm B},\right)^T.
    \end{gathered}
\end{eqnarray}

In the following, we have summarized the various interactions and hopping integrals in matrices that arise in our largest TB model: 

\begin{equation} \label{matrx1}
    \hspace{-0.08\textwidth}
    \gamma(\mathbf{k})=
    \left(
    \begin{array}{c||cccccc|c|c}
        \mu & {\rm A}_s & {\rm A}_p &  {\rm A'}_s & {\rm A'}_p & {\rm B}_s & {\rm B}_p & {\rm B}_d & {\rm B}_s^* \\
        \toprule
        {\rm A}_s &&&&&&&\gamma^{\rm AB}_{sd}[1\times5]&\gamma_{s^*s}^{\rm AB}\\
        {\rm A}_p &&&&&&&\gamma^{\rm AB}_{pd}[3\times5]&(\gamma_{s^*p}^{\rm AB})^T\\
        {\rm A'}_s && \multicolumn{3}{c}{\gamma_{sp}(\mathbf{k})[12\times12]} &&&(\gamma^{\rm AB}_{sd})^*[1\times5]&(\gamma_{s^*s}^{\rm AB})^*\\
        {\rm A'}_p &&&&&&&(-\gamma^{\rm AB}_{pd})^*[3\times5]&(-\gamma_{s^*p}^{\rm AB})^\dagger\\
        {\rm B}_s &&&&&&&\gamma^{\rm BB}_{sd}[1\times5]&\gamma_{s^*s}^{\rm BB} \\
        {\rm B}_p &&&&&&&\gamma^{\rm BB}_{pd}[3\times5]&(\gamma_{s^*p}^{\rm BB})^T \\
        \hline
        & &&&&&&& \\
        {\rm B}_d & (\gamma^{\rm AB}_{sd})^\dagger & (\gamma^{\rm AB}_{pd})^\dagger & (\gamma^{\rm AB}_{sd})^T & (-\gamma^{\rm AB}_{pd})^T & (\gamma^{\rm BB}_{sd})^\dagger & (\gamma^{\rm BB}_{pd})^\dagger & \gamma^{\rm BB}_{dd} & \gamma_{s^*d}^{\rm BB} \\
        & [5\times1] & [5\times3] & [5\times1] & [5\times3] & [5\times1] & [5\times3] & [5\times5] & [1\times5]\\
        & &&&&&&& \\
        \hline
        {\rm B}_s^*& (\gamma_{s^*s}^{\rm AB})^* & (\gamma_{s^*p}^{\rm AB})^*  & \gamma_{s^*s}^{\rm AB} & -\gamma_{s^*p}^{\rm AB} & (\gamma_{s^*s}^{\rm BB})^* & (\gamma_{s^*p}^{\rm BB})^\dagger & (\gamma_{s^*d}^{\rm BB})^\dagger & (\gamma_{s^*s^*}^{\rm BB})
    \end{array}
    \right),
\end{equation}

\begin{equation} \label{matrx2}
    \hspace{-0.08\textwidth}
    \gamma_{sp}(\mathbf{k})=
    \left(
    \begin{array}{c||c|ccc|c|ccc|c|ccc}
        \mu & {\rm A}_s & {\rm A}_x & {\rm A}_y & {\rm A}_z &  {\rm A'}_s & {\rm A'}_x & {\rm A'}_y & {\rm A'}_z & {\rm B}_s & {\rm B}_{px} & {\rm B}_{py} & {\rm B}_{pz} \\
        \toprule
        {\rm A}_s & \gamma^{\rm AA}_{ss} & & \gamma^{\rm AA}_{sp} & & \gamma^{\rm AA'}_{ss} & & \gamma^{\rm AA'}_{sp} & & \gamma^{\rm AB}_{ss} &  & \gamma^{\rm AB}_{sp} & \\
        \hline
        {\rm A}_x &  & & & & & & & & & & &\\
        {\rm A}_y & (\gamma^{\rm AA}_{sp})^\dagger & & \gamma^{\rm AA}_{pp} & & (-\gamma^{\rm AA'}_{sp})^T & & \gamma^{\rm AA'}_{pp} & & (\gamma^{\rm BA}_{sp})^T & & \gamma^{\rm AB}_{pp} &\\
        {\rm A}_z & & & & & & & & & & & &\\
        \hline
        {\rm A'}_s & \gamma^{\rm AA'}_{ss} & & (-\gamma^{\rm AA'}_{sp})^* & & \gamma^{\rm AA}_{ss} & & \gamma^{\rm AA}_{sp} & & (\gamma^{\rm AB}_{ss})^* & & -(\gamma^{\rm AB}_{sp})^*  &\\
        \hline
        {\rm A'}_x & & & & & & & & & & & &\\
        {\rm A'}_y & (\gamma^{\rm AA'}_{sp})^\dagger & & \gamma^{\rm AA'}_{pp} & & (\gamma^{\rm AA}_{sp})^\dagger & & \gamma^{\rm AA}_{pp} & & (-\gamma^{\rm BA}_{sp})^\dagger & & (\gamma^{\rm AB}_{pp})^* &\\
        {\rm A'}_z & & & & & & & & & & & &\\
        \hline
        {\rm B}_s & (\gamma^{\rm AB}_{ss})^* & & (\gamma^{\rm BA}_{sp})^* & & \gamma^{\rm AB}_{ss} & & -\gamma^{\rm BA}_{sp} & & \gamma^{\rm BB}_{ss} &  & \gamma^{\rm BB}_{sp} & \\
        \hline
        {\rm B}_{px} &  & & & &  & & & &   &  &  &  \\
        {\rm B}_{py} & (\gamma^{\rm AB}_{sp})^\dagger & & (\gamma^{\rm AB}_{pp})^* & & -(\gamma^{\rm AB}_{sp})^T & & \gamma^{\rm AB}_{pp} & & (\gamma^{\rm BB}_{sp})^\dagger & & \gamma^{\rm BB}_{pp}&  \\
        {\rm B}_{pz} & & & & &  & & & &   & & &\\
    \end{array}
    \right).
\end{equation}

The interaction parameters defined in Eqs. (\ref{matrx1}) and (\ref{matrx2}) are given by
\begin{equation}\label{eq:gamma_AA'ss}
    \gamma^{\rm A A'}_{ss}=\gamma^{\rm A' A}_{ss}=2\sigma_{ss}^{\text{AA}}(\cos\frac{k_x}{2}+\cos\frac{k_y}{2}+\cos\frac{k_z}{2}),
\end{equation}
\begin{equation}\label{eq:gamma_AAss}
    \gamma^{\rm A A}_{ss}=\gamma^{\rm A' A'}_{ss}=4\tilde{\sigma}_{ss}^{\text{AA}}(\cos\frac{k_x}{2}\cos\frac{k_y}{2}+
    \cos\frac{k_x}{2}\cos\frac{k_z}{2}+\cos\frac{k_y}{2}\cos\frac{k_z}{2} ),
\end{equation}
\begin{equation}\label{eq:gamma_BBss}
    \gamma^{\rm BB}_{ss}=4\sigma_{ss}^{\text{BB}}(\cos\frac{k_x}{2}\cos\frac{k_y}{2}+
    \cos\frac{k_x}{2}\cos\frac{k_z}{2}+\cos\frac{k_y}{2}\cos\frac{k_z}{2}),
\end{equation}
\begin{equation}\label{eq:gamma_ABss}
    \begin{split}
        \gamma^{\rm AB}_{ss}=(\gamma^{\rm A'B}_{ss})^*=4\sigma_{ss}^{\text{AB}}(\cos\frac{k_x}{4}\cos\frac{k_y}{4}\cos\frac{k_z}{4}-i\sin\frac{k_x}{4}\sin\frac{k_y}{4}\sin\frac{k_z}{4}),
    \end{split}
\end{equation}
\begin{equation}\label{eq:gamma_BBpp}
    \begin{split}
        \frac{\gamma^{\rm BB}_{pp}}{2}
        =t^{p_ip_j}(\boldsymbol{\delta}_1^+)\cos(\frac{k_y}{2}+\frac{k_z}{2})&+t^{p_ip_j}(\boldsymbol{\delta}_4^+)\cos(\frac{k_y}{2}-\frac{k_z}{2}) +t^{p_ip_j}(\boldsymbol{\delta}_2^+)\cos(\frac{k_x}{2}+\frac{k_z}{2})+t^{p_ip_j}(\boldsymbol{\delta}_5^+)\cos(\frac{k_x}{2}-\frac{k_z}{2}) \\
        &+t^{p_ip_j}(\boldsymbol{\delta}_3^+)\cos(\frac{k_x}{2}+\frac{k_y}{2})+t^{p_ip_j}(\boldsymbol{\delta}_6^+)\cos(\frac{k_x}{2}-\frac{k_y}{2}),\\
    \end{split}
\end{equation}
where
\begin{equation}\label{eq:gamma_BBpp_matrix}
    \begin{split}
        &t^{p_ip_j}(\boldsymbol{\delta}_1^+)=
        \begin{pmatrix}
        \pi_{pp}^{\rm B} & 0 & 0 \\
            0 & p_+ & p_- \\
            0 & p_- & p_+
        \end{pmatrix}
        \hspace{0.7cm}
        t^{p_ip_j}(\boldsymbol{\delta}_2^+)=
        \begin{pmatrix}
            p_+ & 0 & p_- \\
            0 & \pi_{pp}^{\rm B} & 0 \\
            p_- & 0 & p_+
        \end{pmatrix}
        \hspace{0.7cm}
        t^{p_ip_j}(\boldsymbol{\delta}_3^+)=
        \begin{pmatrix}
            p_+ & p_- & 0 \\
            p_- & p_+ & 0 \\
            0 & 0 & \pi_{pp}^{\rm B}
        \end{pmatrix}
        \\
        &t^{p_ip_j}(\boldsymbol{\delta}_4^+)=
        \begin{pmatrix}
            \pi_{pp}^{\rm B} & 0 & 0 \\
            0 & p_+ & -p_- \\
            0 & -p_- & p_+
        \end{pmatrix}
        \hspace{0.1cm}
        t^{p_ip_j}(\boldsymbol{\delta}_5^+)=
        \begin{pmatrix}
            p_+ & 0 & -p_- \\
            0 & \pi_{pp}^{\rm B} & 0 \\
            -p_- & 0 & p_+
        \end{pmatrix}
        \hspace{0.1cm}
        t^{p_ip_j}(\boldsymbol{\delta}_6^+)=
        \begin{pmatrix}
            p_+ & -p_- & 0 \\
            -p_- & p_+ & 0 \\
            0 & 0 & \pi_{pp}^{\rm B}
        \end{pmatrix}.
    \end{split}
\end{equation}
Here, $p_{\pm}\equiv (\sigma_{pp}^{\rm B}\pm\pi_{pp}^{\rm B})/2$ where $\sigma_{pp}^B$ and $\pi_{pp}^B$ represent the $\sigma$ and $\pi$ bonds for the $p$ orbitals of the $\rm B$ atoms. The elements $t^{p_ip_j}(\boldsymbol{\delta})[i,j]$ of (\ref{eq:gamma_BBpp_matrix}) represent the coupling between orbitals $p_i$ and $p_j$ in the direction $\boldsymbol{\delta}$. The equivalent expressions for the $\rm A$ elements are obtained as
\begin{equation}\label{eq:gamma_AApp}
    \frac{\gamma^{\rm AA'}_{pp}}{2}=\frac{\gamma^{\rm A'A}_{pp}}{2}=
    \begin{pmatrix}
        \sigma_{pp}^{\text{A}} & 0 & 0 \\
            0 & \pi_{pp}^{\text{A}} & 0 \\
            0 & 0 & \pi_{pp}^{\text{A}}
    \end{pmatrix}\cos\frac{k_x}{2}+
    \begin{pmatrix}
        \pi_{pp}^{\text{A}} & 0 & 0 \\
            0 & \sigma_{pp}^{\text{A}} & 0 \\
            0 & 0 & \pi_{pp}^{\text{A}}
    \end{pmatrix}\cos k_y+
    \begin{pmatrix}
        \pi_{pp}^{\text{A}} & 0 & 0 \\
            0 & \pi_{pp}^{\text{A}} & 0 \\
            0 & 0 & \sigma_{pp}^{\text{A}}
    \end{pmatrix}\cos k_z.
\end{equation}
\begin{equation}\label{eq:gamma_AApp_N}
    \begin{split}
       \gamma^{\rm AA}_{pp}=\gamma^{\rm A'A'}_{pp}=\gamma^{\rm BB}_{pp}[\sigma_{pp}^{\rm B}\leftarrow\tilde{\sigma}_{pp}^{\rm A}, \pi_{pp}^{\rm B}\leftarrow\tilde{\pi}_{pp}^{\rm A}],
    \end{split}
\end{equation}
where $\gamma^{\rm BB}_{pp}$ is given by (\ref{eq:gamma_BBpp}) with a new set of parameters, i.e., $\tilde{\sigma}_{pp}^{\rm A}$ and $\tilde{\pi}_{pp}^{\rm A}$.
For the cross couplings AB, the interactions are given by
\begin{equation}\label{eq:gamma_ABpp}
    \hspace{-0.5cm}
    \begin{split}
        &\frac{\gamma^{\rm AB}_{pp}}{4}=\frac{(\gamma^{\rm A'B}_{pp})^*}{4}=
        p_{2}^{+}
        \begin{pmatrix}
            1 & 0 & 0 \\
            0 & 1 & 0 \\
            0 & 0 & 1 
        \end{pmatrix}
        \left(\cos\frac{k_x}{4}\cos\frac{k_y}{4}\cos\frac{k_z}{4}-i\sin\frac{k_x}{4}\sin\frac{k_y}{4}\sin\frac{k_z}{4}\right)
        -
        p_{2}^{-}
        \begin{pmatrix}
            0 & 1 & 0 \\
            1 & 0 & 0 \\
            0 & 0 & 0 
        \end{pmatrix}
        \\&\times\left(\sin\frac{k_x}{4}\sin\frac{k_y}{4}\cos\frac{k_z}{4}-i\cos\frac{k_x}{4}\cos\frac{k_y}{4}\sin\frac{k_z}{4}\right) 
         -
        p_{2}^{-}
        \begin{pmatrix}
            0 & 0 & 1 \\
            0 & 0 & 0 \\
            1 & 0 & 0 
        \end{pmatrix}
        \left(\sin\frac{k_x}{4}\cos\frac{k_y}{4}\sin\frac{k_z}{4}-i\cos\frac{k_x}{4}\sin\frac{k_y}{4}\cos\frac{k_z}{4}\right) \\
        &-
        p_{2}^{-}
        \begin{pmatrix}
            0 & 0 & 0 \\
            0 & 0 & 1 \\
            0 & 1 & 0 
        \end{pmatrix}
        \left(\cos\frac{k_x}{4}\sin\frac{k_y}{4}\sin\frac{k_z}{4}-i\sin\frac{k_x}{4}\cos\frac{k_y}{4}\cos\frac{k_z}{4}\right),
    \end{split}
\end{equation}
where $p_{2}^{-}\equiv (\sigma_{pp}^{\text{AB}}-\sigma_{pp}^{\text{AB}})/3$ and $p_{2}^{+}\equiv (\sigma_{pp}^{\text{AB}}+2\sigma_{pp}^{\text{AB}})/3$. Also,

\begin{equation}\label{eq:gamma_BBsp}
    \gamma^{\rm BB}_{sp}=2\sqrt{2}i\sigma_{sp}^{\text{BB}}
\left(
        \sin\frac{k_x}{2}(\cos\frac{k_y}{2}+\cos\frac{k_z}{2} ),
        \sin\frac{k_y}{2}(\cos\frac{k_z}{2}+\cos\frac{k_x}{2} ),
        \sin\frac{k_z}{2}(\cos\frac{k_x}{2}+\cos\frac{k_y}{2} )
    \right),
\end{equation}
\begin{equation}\label{eq:gamma_AAsp}
    \gamma^{\rm AA'}_{sp}=\gamma^{\rm A'A}_{sp}=2i\sigma_{sp}^{\rm AA}\left(\sin\frac{k_x}{2},\sin\frac{k_y}{2},\sin\frac{k_z}{2}\right),
\end{equation}

\begin{equation}\label{eq:gamma_AApp_Nn}
    \begin{split}
       \gamma^{\rm AA}_{sp}=\gamma^{\rm A'A'}_{sp}=\gamma^{\rm BB}_{sp}[\sigma_{sp}^{\rm B}\leftarrow\tilde{\sigma}_{sp}^{\rm A}],
    \end{split}
\end{equation}
\begin{equation}\label{eq:gamma_ABsp}
\begin{gathered}
    \gamma^{\rm AB}_{sp}=-(\gamma^{\rm A'B}_{sp})^*=-\frac{4\sigma_{sp}^{\rm AB}}{\sqrt{3}} 
        \left(\cos\frac{k_x}{4}\sin\frac{k_y}{4}\sin\frac{k_z}{4}-  i\sin\frac{k_x}{4}\cos\frac{k_y}{4}\cos\frac{k_z}{4}, 
        \sin\frac{k_x}{4}\cos\frac{k_y}{4}\sin\frac{k_z}{4}-i\cos\frac{k_x}{4}\sin\frac{k_y}{4}\cos\frac{k_z}{4}, \right. \\ \left.
        \sin\frac{k_x}{4}\sin\frac{k_y}{4}\cos\frac{k_z}{4}-i\cos\frac{k_x}{4}\cos\frac{k_y}{4}\sin\frac{k_z}{4}
\right),
\end{gathered}
\end{equation}
\begin{equation}\label{eq:gamma_BAsp}
    \gamma^{\rm BA}_{sp}=-(\gamma^{\rm BA'}_{sp})^*=\frac{\sigma_{sp}^{\rm BA}}{\sigma_{sp}^{\rm AB}}\gamma^{\rm AB}_{sp}.
\end{equation}
The interaction of the $d$ orbitals with those of $s,p,d$ are summarized as follows.
\begin{equation}\label{eq:gamma_dd}
    \gamma_{dd}=\sum_{\boldsymbol{\delta}\in \boldsymbol{\delta_{\rm BB}^+}}2t_{dd}(\boldsymbol{\delta})\cos(\mathbf{k\cdot \boldsymbol{\delta}}),
\end{equation}
\begin{equation}\label{eq:SK_dd}
    \hspace{-2cm}
    \begin{split}
        t_{dd}(\boldsymbol{\delta})&=
        \begin{pmatrix}
            3l^2m^2 & 3lm^2n & 3l^2mn & \frac{3}{2}lm(l^2-m^2) & \sqrt{3}lm[n^2-\frac{1}{2}(l^2+m^2)] \\
            & 3m^2n^2 & 3lmn^2 & \frac{3}{2}mn(l^2-m^2) & \sqrt{3}mn[n^2-\frac{1}{2}(l^2+m^2)] \\
            && 3 l^2n^2 & \frac{3}{2}nl(l^2-m^2) & \sqrt{3}nl[n^2-\frac{1}{2}(l^2+m^2)] \\
            &\text{T}&& \frac{3}{2}(l^2-m^2)^2 & \sqrt{3}(l^2-m^2)[n^2-\frac{1}{2}(l^2+m^2)]  \\
            &&&& [n^2-\frac{1}{2}(l^2+m^2)]^2
        \end{pmatrix}
        \sigma_{dd} \\
        &+
        \begin{pmatrix}
           l^2+m^2-4l^2m^2 & ln(1-4m^2) & mn(1-4l^2) & 2lm(m^2-l^2) & -2\sqrt{3}lmn^2 \\
           & m^2+n^2-4m^2n^2 & lm(1-4n^2) & -mn[1+2(l^2-m^2)] & \sqrt{3}mn(l^2+m^2-n^2) \\
           && l^2+n^2-4l^2n^2 & nl[1-2(l^2-m^2)] &\sqrt{3}ln(l^2+m^2-n^2) \\
          \hspace{1.2cm}\text{T} &&& l^2+m^2-(l^2-m^2)^2 & \sqrt{3}n^2(m^2-l^2) \\
           &&&& 3n^2(l^2+m^2)
        \end{pmatrix}
        \pi_{dd} \\
        &+
        \begin{pmatrix}
            n^2+l^2m^2 & ln(m^2-1) & mn(l^2-1) & \frac{1}{2}lm(l^2-m^2) & \frac{\sqrt{3}}{2}lm(1+n^2) \\
            & l^2+m^2n^2 & lm(n^2-1) & mn(1+\frac{1}{2}(l^2-m^2)) & -\frac{\sqrt{3}}{2}mn(l^2+m^2) \\
            &  & m^2+l^2n^2 & -nl(1-\frac{1}{2}(l^2-m^2)) & -\frac{\sqrt{3}}{2}ln(l^2+m^2) \\
            & \text{T} &  & n^2+\frac{1}{4}(l^2-m^2)^2 & \frac{\sqrt{3}}{4}(1+n^2)(l^2-m^2) \\
            &  &  &  & \frac{3}{4}(l^2+m^2)^2
        \end{pmatrix}
        \delta_{dd}, 
    \end{split}
\end{equation}
in which $\sigma_{dd}$, $\pi_{dd}$ and $\delta_{dd}$ are the hopping parameters, and T denotes the transpose of the upper-triangular matrix. $\boldsymbol{\delta}_{BB}^{+}$ is the half of the BB vectors presented in Table \ref{tab_vc} with a positive sign. Also, $\{l,m,n\}$ are respectively the $x$, $y$ and $z$ components of the direction of $\boldsymbol{\delta}$. 
\begin{equation}\label{eq:gamma_BBpd}
    \gamma^{\rm BB}_{pd}=\sum_{\boldsymbol{\delta}\in \boldsymbol{\delta_{BB}^+}}2it_{pd}(\boldsymbol{\delta})\sin(\mathbf{k\cdot \boldsymbol{\delta}}), 
    \end{equation}
where $t_{pd}(\boldsymbol{\delta})$ is the $p$-$d$ interaction matrix given by 
\begin{equation}\label{eq:SK_pd}
    \begin{split}
        t_{pd}(\boldsymbol{\delta})&=
        \begin{pmatrix}
            \sqrt{3}l^2m & \sqrt{3}lmn & \sqrt{3}l^2n & \frac{\sqrt{3}}{2}l(l^2-m^2) & l[n^2-\frac{1}{2}(l^2+m^2)]\\
            \sqrt{3}lm^2 & \sqrt{3}m^2n & \sqrt{3}lmn & \frac{\sqrt{3}}{2}m(l^2-m^2) & m[n^2-\frac{1}{2}(l^2+m^2)]\\
            \sqrt{3}lmn & \sqrt{3}mn^2 & \sqrt{3}ln^2 & \frac{\sqrt{3}}{2}n(l^2-m^2) & n[n^2-\frac{1}{2}(l^2+m^2)]
        \end{pmatrix}
        \sigma_{pd} \\
        &+
        \begin{pmatrix}
           m(1-2l^2) & -2lmn & n(1-2l^2) & l(1-l^2+m^2) & -\sqrt{3}ln^2 \\
           l(1-2m^2) & m(1-2m^2) & -2lmn & -m(1+l^2-m^2) & -\sqrt{3}mn^2 \\
           -2lmn & m(1-2n^2) & l(1-2n^2) & -n(l^2+m^2) & \sqrt{3}n(l^2+m^2)
        \end{pmatrix}
        \pi_{pd}.
    \end{split}
\end{equation}
For the $\gamma^{\rm AB}_{pd}$ interaction, no significant simplifications are available, and it is most convenient to use Eq. (\ref{eq:gam}) directly,
\begin{equation}\label{eq:gamma_ABpd}
\gamma^{\rm AB}_{pd}=\sum_{\boldsymbol{\delta}\in \boldsymbol{\delta_{AB}}}t_{pd}(\boldsymbol{\delta})\exp({i \mathbf{k}\cdot \boldsymbol{\delta}})
\end{equation}
Finally, the interaction between the $s$ and $d$ orbitals can be expressed by 
\begin{equation}\label{eq:gamma_BBsd}
    \gamma_{sd}^{\rm BB}=\sum_{\boldsymbol{\delta}\in \boldsymbol{\delta_{\rm BB}^+}}2t_{sd}(\boldsymbol{\delta})\cos(\mathbf{k\cdot \boldsymbol{\delta}}), 
\end{equation}
\begin{equation}\label{eq:SK_sd}
    t_{sd}^{\rm BB}(\boldsymbol{\delta})=\sigma_{sd}^{\rm BB}\left(\sqrt{3}lm, \sqrt{3}mn, \sqrt{3}ln, \frac{\sqrt{3}}{2}(l^2-m^2), n^2-\frac{1}{2}(l^2+m^2)\right),
\end{equation}
and $\gamma^{\rm AB}_{sd}$ is found to be 
\begin{equation}\label{eq:gamma_ABsd}
    \gamma^{\rm AB}_{sd}=(\gamma^{\rm A'B}_{sd})^*=\frac{\sigma_{sd}^{\rm AB}}{\sigma_{sp}^{\rm AB}}\left(\gamma^{\rm AB}_{sp_z},\gamma^{\rm AB}_{sp_x},\gamma^{\rm AB}_{sp_y},0 , 0 \right).
\end{equation}
A virtual $s^*$ orbital may be introduced into the formulations to represent the 4s/5s/6s/7s orbitals of Si/Ge/Sn/Pb in order to obtain better fittings with less deviations with respect to the DFT band structure. The interaction integrals, corresponding to $s^*$, can be calculated using 
\begin{equation}\label{eq:gamma_s*}
    \gamma_{s^*\mu}^{\rm B\alpha}=\frac{\sigma_{s^*\mu}^{\rm B\alpha}}{\sigma_{s\mu}^{\rm B\alpha}}\gamma_{s\mu}^{\rm B\alpha}, \hspace{0.5cm}
    \gamma_{s^*s^*}^{\rm BB}=\frac{\sigma_{s^*s^*}^{\rm BB}}{\sigma_{ss}^{\rm BB}} \gamma_{ss}^{\rm BB} , \hspace{0.5cm} \alpha=\{{\rm A,A',B}\},\mu=\{s,p,d\}.
\end{equation}

\begin{figure}[t!]
\includegraphics[width=0.35\textwidth]{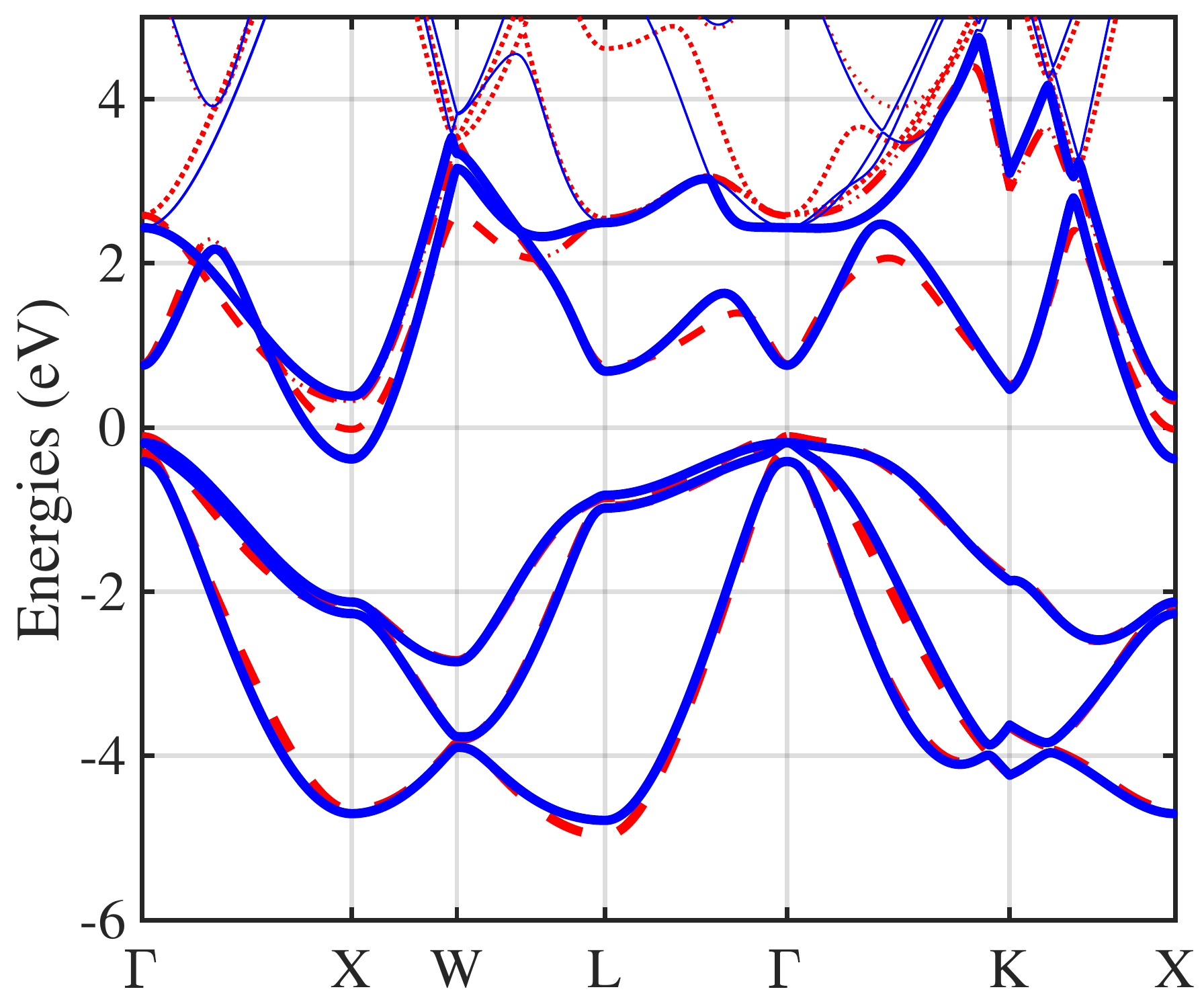}
\caption{\label{bs_18x} The band structure of $\rm Mg_2Ge$ along the high-symmetry path $\rm \Gamma XWL\Gamma KX$. The strain is set to zero and the Fermi level is shifted to $E=0$. The solid curves belong to the machine-learned eighteen-band TB model whereas the dashed curves are obtained by DFT.}
\end{figure}

Also, we may summarize the $p$ orbital matrix elements for the SOI operator $\eta_{\mu\nu,\sigma\sigma'}$ as
\begin{equation}\label{eq:gamma_SOC}
    \eta_{p_ip_j\sigma\sigma'}=
    \frac{\eta_\text{SO}}{2}\left(
    \begin{array}{c||cccccc}
         p_{i\sigma} & p_{x\uparrow} & p_{x\downarrow} & p_{y\uparrow} & p_{y\downarrow} & p_{z\uparrow} & p_{z\downarrow} \\
         \toprule
        p_{x\uparrow}   & 0 & 0 & -i &0  &0   & 1\\
        p_{x\downarrow} & 0 & 0 & 0  & i & -1 & 0 \\
        p_{y\uparrow}   &i  &0  &0 &0 &0 & -i \\
        p_{y\downarrow} &0  & -i  &0 & 0 &-i & 0\\
        p_{z\uparrow}   &0  &-1  &0 &i &0 &0\\
        p_{z\downarrow} &1 &0 &i &0 &0 &0
    \end{array}
    \right).
\end{equation}

\section{The total band structure using the machine-learned eighteen-band TB model}\label{apx:bs}

As pointed out in the main text, the deviation of the conduction bands in our machine-learned five-band TB model originates from the truncation of the more complete basis set given by Eq.~(\ref{eq:18xbasis}). A full conduction band is a complicated hybridization of different excited orbitals such as the $d$ orbitals in $\rm Mg_2X$. To show how the inclusion of higher excited states in our TB models can improve the fitting of conduction bands in the five-band TB model, we have employed the largest TB model we constructed, i.e., the eighteen-band TB model, and plotted its results in Fig.~\ref{bs_18x}. We have considered a representative case where strain is set to zero and $\rm X = Ge$. The dashed red curves are the DFT band structure whereas the solid blue curves are the band structure from our machine-learned eighteen-band TB model. Compared to Fig.~\ref{bs_5x}(b2), we clearly see that the conduction bands are now greatly improved and more delicate features are captured by the eighteen-band TB model. The same improvement is accessible through the eighteen-band TB model when applied to the cases shown in Fig.~\ref{bs_5x}. 

\section{Machine-learning obtained parameters to the five-band TB model}\label{apx:cs}
Table~\ref{tab:TBparam} summarizes the optimized bonding parameter values obtained for the five-band TB model, presented in the main text, to describe $\rm Mg_2X (X=Si,Ge,Sn,Pb)$. We have included the obtained parameter values when strain is $\varepsilon=0\%,\pm 10\%$. The presented bonding parameter values, on-site energies, and spin-orbit coupling strength $\eta_\text{SO}$ in Table~\ref{tab:TBparam} reproduce the valence band structure within the entire Brillouin zone. 

\begin{table}[h]
\label{tab:TBparam}
\makebox[\textwidth][c]{
\begin{tabular}{|cc||cc|cc|c|cc|c|}
\hline
\hline
                &type  & \multicolumn{2}{c}{AA} &  \multicolumn{2}{c}{BB}                  & AB        & \multicolumn{2}{c}{on-site}       & SOC       \\
                \hline
strain          &      & $\sigma_{ss}$ (eV)   & $\tilde{\sigma}_{ss}$ (eV) & $\sigma_{pp}$ (eV) & $\pi_{pp}$ (eV)     & $\sigma_{sp}$ (eV) & $E_{\text{A}s}$ (eV)   & $E_{\text{B}p}$ (eV)    & $\eta_\text{SO}$ (eV)   \\
\hline
                &Si  & 0.0400    & 0.1036    & 0.6370    & 0.0002 & 0.6559    & 1.9654  & -2.6307 & 0.0190  \\
$\varepsilon=-10\%$&Ge  & 0.0793    & 0.1092    & 0.6251    & 0.0039 & 0.7674    & 2.0422  & -2.6611 & 0.1234  \\
                &Sn  & 0.0327    & 0.0955    & 0.6451    & 0.0147 & 0.6695    & 1.5619  & -2.8796 & 0.3059  \\
                &Pb  & -0.0955   & 0.0932    & 0.5718    & 0.0344 & 0.9046    & 0.9047  & -3.0787 & 1.0731  \\
\hline
                &Si  & -0.0759   & 0.0660    & 0.4952    & 0.0052 & 0.6642    & 1.3375  & -2.2480 & 0.0190  \\
$\varepsilon=0$    &Ge  & -0.1499   & 0.0587    & 0.4362    & 0.0233 & 0.8279    & 1.0853  & -2.1530 & 0.1220  \\
                &Sn  & -0.0762   & 0.0678    & 0.4430    & 0.0281 & 0.7744    & 0.8680  & -2.1290 & 0.2836  \\
                &Pb  & -0.1913   & 0.0479    & 0.3936    & 0.0451 & 0.8707    & 0.4543  & -2.0863 & 0.9174  \\
\hline
                &Si  & -0.1600   & 0.0454    & 0.3044    & 0.0304 & 0.7843    & 0.6505  & -1.5692 & 0.0186  \\
$\varepsilon=+10\%$&Ge  & -0.2273   & 0.0331    & 0.2730    & 0.0425 & 0.8346    & 0.4864  & -1.4543 & 0.1135  \\
                &Sn  & -0.1388   & 0.0590    & 0.1994    & 0.0632 & 0.8655    & 0.1442  & -1.2072 & 0.2777  \\
                &Pb  & -0.2220   & 0.0400    & 0.1733    & 0.0738 & 0.8700    & -0.1516 & -1.1477 & 0.8055 \\
\hline
\hline
\end{tabular}

}
\caption{Hopping parameters, on-site energies, and spin-orbit coupling strength for the five-band TB model.}
\end{table}

\twocolumngrid

\end{document}